\documentclass[journal,12pt,draftclsnofoot,onecolumn]{IEEEtran}
\usepackage{color}
\usepackage{balance}
\usepackage{graphicx}
\usepackage{enumerate}
\usepackage{subfigure}
\usepackage{algorithm}
\usepackage{algorithmic}
\usepackage{booktabs}
\usepackage{cite}
\usepackage{amssymb}
\usepackage{amsmath}
\usepackage{float}
\usepackage{float}
\usepackage{bm}

\begin{document}

\title{Intelligent Reflecting Surface Assisted Anti-Jamming Communications: A Fast Reinforcement Learning Approach
}
\title{Privacy-Preserving Federated Learning for UAV-Enabled Networks: Learning-Based Joint Scheduling and Resource Management}

\author{

  Helin~Yang,~\IEEEmembership{Member,~IEEE},
  Jun~Zhao,~\IEEEmembership{Member,~IEEE},
  Zehui~Xiong,~\IEEEmembership{Member,~IEEE},
  Kwok-Yan~Lam,~\IEEEmembership{Senior Member,~IEEE},
  Sumei~Sun, ~\IEEEmembership{Fellow,~IEEE},
  and Liang~Xiao,~\IEEEmembership{Senior Member,~IEEE}

\thanks{\small{H. Yang,  Z. Xiong, J. Zhao, and K. Y. Lam are with the School of Computer Science and Engineering, and Strategic Centre for Research in Privacy-Preserving Technologies, Nanyang Technological University, Singapore 639798 (e-mail: hyang013@e.ntu.edu.sg,  zxiong002@e.ntu.edu.sg, junzhao@ntu.edu.sg, kwokyan.lam@ntu.edu.sg).}}

\thanks{ \small{S. Sun is with the Institute for Infocomm Research, Agency for Science, Technology, and Research, Singapore 138632 (e-mail: sunsm@i2r.a-star.edu.sg).}}

\thanks{ \small{L. Xiao  is with the Department of Information and Communication Engineering, Xiamen University, Xiamen 361005, China (e-mail: lxiao@xmu.edu.cn).}}

}



\maketitle

\begin{abstract}

Unmanned aerial vehicles (UAVs) are capable of serving as flying base stations (BSs) for supporting data collection, artificial intelligence (AI) model training, and wireless communications. However, due to the privacy concerns of devices and limited computation or communication resource of UAVs, it is impractical to send raw data of devices to UAV servers for model training. Moreover, due to the dynamic channel condition and heterogeneous computing capacity of devices in UAV-enabled networks, the reliability and efficiency of data sharing require to be further improved.  In this paper, we develop an asynchronous federated learning (AFL) framework for multi-UAV-enabled networks, which can provide asynchronous distributed computing by enabling model training locally without transmitting  raw sensitive data to UAV servers. The device selection strategy is also introduced into the AFL framework to keep the low-quality devices from affecting the learning efficiency and accuracy. Moreover, we propose an asynchronous advantage actor-critic (A3C) based joint device selection, UAVs placement, and resource management algorithm to enhance the federated convergence speed and accuracy. Simulation results demonstrate that our proposed framework and algorithm achieve higher learning accuracy and faster federated execution time compared to other existing solutions.  \vspace{10pt}

\textbf{\emph{Index Terms}}---Unmanned aerial vehicle, data sharing, asynchronous federated learning, scheduling, resource management, asynchronous advantage actor-critic.
\end{abstract}

\IEEEpeerreviewmaketitle
\section{Introduction}

\IEEEPARstart{W}{ith} the rapid deployment of fifth-generation (5G) and beyond wireless networks, unmanned aerial vehicles (UAVs) have been emerged as a promising candidate paradigm to provide communication and computation services for ground mobile devices in sport stadiums, outdoor events, hot spots, and remote areas [1]-[3], where aerial base stations (BSs) or servers can be mounted on UAVs. Given the benefits of agility, flexibility, mobility, and beneficial line-of-sight (LoS) propagation, UAV-enabled networks have been widely applied in quick-response wireless communications, data collection, artificial intelligence (AI) model training, and coverage enhancement [1]. Correspondingly, UAV-assisted computation and communication tasks have attracted significant interest recently in 5G and beyond wireless networks [3].

Again, deploying UAVs as flight BSs is able to flexibly to collect data and implement AI model training for ground devices, but it is impractical for a large number of devices to transmit their  raw data to UAV servers due to the privacy concern and limited communication resources for data transmission [4]. Moreover, as the energy capacity, storage, and computational ability of UAVs are limited, it is still challenging for UAVs to process/training a large amount of  raw data [1], [2], [4]. In face of the challenges, federated learning (FL) [5], [6] emerges as a promising paradigm aiming to protect device privacy by enabling devices to train AI model locally without sending their raw data to a server. Instead of training AI model at the data server, FL enables devices to execute local training on their own data, which generally uses the gradient descent optimization method [7].

By using  FL, UAVs can perform distributed AI tasks for ground mobile devices without relying on any centralized BS, and devices also do not need to send any raw data to UAVs during the training [4]. In particular, wireless devices use their respective local datasets to train AI models, and upload the local model parameters to an FL UAV server for model aggregation. After collecting the local model parameters from devices, the UAV server then aggregates the updated model parameters before broadcasting the parameters to associated devices for another round of local model training. During the process, keeping raw data at devices not only preserves privacy but also reduces network traffic congestion. A number of rounds are performed until a target learning accuracy is obtained. In essence, FL allows UAV-enabled wireless networks to train AI models in an efficient way, compared with centralized cloud-centric frameworks. However, since the parameters of AI models in FL need to be exchanged between UAV servers and devices, the FL convergence and task consensus for UAV servers will inevitably be affected by transmission latency [4]. In addition, due to the mobility of UAVs and devices, dynamic channel conditions can negatively affect the FL convergence [8]. Moreover, UAV servers are generally limited in terms of computation and energy resources, resulting in the necessity of design for efficient scheduling and management approaches to minimize the FL execution time [1], [4], [9].  Hence, the aforementioned challenges pertaining to FL specifically that need to be investigated for large scale implementation of efficient FL in UAV-enabled wireless networks.

\subsection{Related Works}

 In recent years, implementing FL in wireless networks has attracted many research efforts, and lots of studies have presented their studies that how to adopt FL to improve the learning efficiency [10]-[17]. Tran $et~al.$ [10] formulated an FL framework over a wireless network as an optimization problem that minimizes the sum of FL aggregation latency and total device energy consumption. In addition, the frequency of global aggregation under different resource constraints (i.e., device central processing unit (CPU), transmission delay, and model accuracy) has been optimized in [11], [12]. However, in [10]-[12], due to the limited wireless spectrum, it is difficult to implement practical wireless FL applications when all mobile devices are involved in each aggregation iteration. Hence, some studies proposed to apply the device selection/scheduling scheme to improve the convergence speed of FL [13]-[16]. The authors in [13] studied the relationship between the number of rounds and different device scheduling policies (i.e., random scheduling, round-robin, and proportional fair), and compared their performances by simulations. A deep reinforcement learning (DRL)-based device selection algorithm was proposed to enhance the reliability and efficiency in an asynchronous federated learning manner [14]. Shi $et~al.$ [9] formulated a joint channel allocation and device scheduling problem to study the convergence speed of FL, but the solution to the problem is not globally optimized. In [15], the authors introduced a novel hierarchical federated edge learning framework, and a joint device and resource scheduling algorithm was proposed to minimize both the system energy and FL execution delay cost. A greedy scheduling scheme was proposed in [16] that manages a part of clients to participate in the global FL aggregation according to their resource conditions. However, the work [16] failed to verify the effectiveness of the scheme in the presence of dynamic channels and computation capabilities.

To the best of our knowledge, there are several studies [4], [8], [9], [18]-[22] that investigated how to apply FL to improve AI model learning efficiency in UAV-enabled communication scenarios. For instance, the authors in [8], [18] developed a novel framework to enable FL within UAV swarms, and a joint power allocation and flying trajectory scheme was proposed to improve the FL learning efficiency [8]. Lim $et~al.$ [19] proposed an FL-based sensing and collaborative learning approach for UAV-enabled internet of vehicles (IoVs), where UAVs collect data and train AI model for IoVs. In addition, Ng $et~al.$ [20] presented the use of UAVs as flight relays to support the wireless communications between IoVs and the FL server, hence enhancing the accuracy of FL. Shiri $et~al.$ [9] adopted FL and mean-field game (MFG) theory to address the online path design problem of massive UAVs, where each UAV can share its own model parameters of AI with other UAVs in a federated manner. The work in [4] provided  discussions of several possible applications of FL in UAV-enabled wireless networks, and introduced the key challenges, open issues, and future research directions therein. To preserve the privacy of devices, a two-stage FL algorithm among the devices, UAVs, and heterogeneous computing platform was proposed to collaboratively predict the content caching placement in a heterogeneous computing architecture [21]. Furthermore, the authors in [22] introduced a blockchain-based collaborative scheme in the proposed secure FL framework for UAV-assisted mobile crowdsensing, which effectively prevents potential security and privacy threats from affecting secure communication performance. However, most of the studies [4], [8], [9], [18]-[22] did not jointly consider the importance of the tasks (i.e., learning accuracy and execution time) of FL that UAVs perform, and the optimal dynamic scheduling and resource management of UAVs to complete the tasks, subject to the resource constraints of UAVs and different computational capabilities of devices.

Joint management of both UAVs trajectory (or UAVs placement) and resource management has been widely studied to optimize the network performance, such as in [23]--[28]. In [23] and [24], a joint optimization for user association, resource allocation, and UAV placement was studied in multi-UAV wireless networks, to guarantee the quality of services (QoS) of mobile devices. However, the conventional optimization methods adopted in [23], [24] may be not effective as the UAVs-enabled environment are dynamic and complex. To address this issue, reinforcement learning (RL) or DRL has been adopted to learn the decision-making strategy [25]-[28]. The authors in [25], [26] applied the multi-agent DRL algorithm to address the joint UAVs trajectory and transmit power optimization problem, and also demonstrated the effectiveness of multi-agent DRL algorithms compared with baseline schemes. Considering the fact that UAVs have continuous action spaces (i.e., trajectory, location, and transmit power), Zhu $et~al.$ [27] proposed an actor-critic (AC)-based algorithm to jointly adjust UAV's transmission control and three -dimensional (3D) flight to maximize the network throughput over continuous action spaces. Similarly, in [28], the authors adopted the AC-based algorithm to optimize the device association, resource allocation, and trajectory of UAVs in dynamic environments. However, to the best of our knowledge, there are no studies applying RL or DRL to achieve the implementation of FL for the UAVs-enabled wireless network, and optimize the FL convergence or accuracy by jointly designing scheduling and resource management.

\subsection{Contributions and Organization}

Motivated by the aforementioned challenges, this paper first develops an asynchronous federated learning (AFL) to achieve fast convergence speed and high learning accuracy for multi-UAV-enabled wireless networks. The proposed framework can enable mobile devices to train their AI models locally and asynchronously upload the model parameters to UAV servers for model aggregation without uploading the raw private data, which reduces communication cost and privacy threats, as well as improves the round efficiency and model accuracy. Considering the dynamic environment and different computational capabilities of devices, we also propose an asynchronous advantage actor-critic (A3C)-based joint device selection, UAVs placement, and resource management algorithm to further enhance the overall federated execution efficiency and model accuracy. The main contributions of this paper are summarized as follows.
    \begin{itemize}
\item We develop a novel privacy-preserving AFL framework for a multi-UAV-enabled wireless network to aggregate and update the AI model parameters in an asynchronous manner, where the mobile devices with high communication and computation capabilities are selected to participate in global aggregation instead of waiting for all associated devices to accomplish their local model update.
\item We propose an A3C-based joint device selection, UAVs placement, and resource management algorithm to further minimize the federated execution time and learning accuracy loss under dynamic environments as well as large-scale continuous action spaces.
\item We conducte extensive simulations to demonstrate that the proposed AFL framework and A3C-based algorithm can significantly improve the FL model accuracy, convergence speed, and federated execution efficiency compared to other baseline solutions under different scenarios.
\end{itemize}

\emph{Organization:} The remainder of this paper is organized as follows. The system model and problem formulation are provided in Section II. Section III proposes the AFL framework and A3C-based algorithm for UAV-enabled wireless networks.  Simulation results and analysis are provided in Section IV, and the paper is concluded in Section V.

\section{System Model and Problem Formulation}



We consider an FL instance consisting of a number of ground
devices, which are associated with several parameter
servers residing at multiple UAVs in the sky. As shown in Fig. 1,
the multi-UAV-enabled network consists of $N$ UAVs
and $K$ single-antenna devices, denoted by
$\mathcal{N} = \{ 1,...,N\} $ and $\mathcal{K} = \{ 1,...,K\} $,
respectively. In practical wireless networks, the ground BS may
face congestion due to a BS malfunction, or a temporary festival
or a big sport event, and even ground BS may fail to provide
coverage in some remote areas. In this case, the deployment of
UAVs has been emerged as a potential technique for providing
wireless coverage for ground devices. As illustrated in Fig. 1,
mobile devices are randomly located on the ground, and multiple
UAVs fly in the sky to provide wireless services for them by
using frequency division multiple access (FDMA). Here, the unity
frequency reuse (UFR) design is employed in multi-UAV networks to
enhance the spectrum utilization, where the communication band is
reused across all cells with each cell being associated with one
UAV [23]-[26].

\begin{figure}
\centering
\includegraphics[width=0.95\columnwidth]{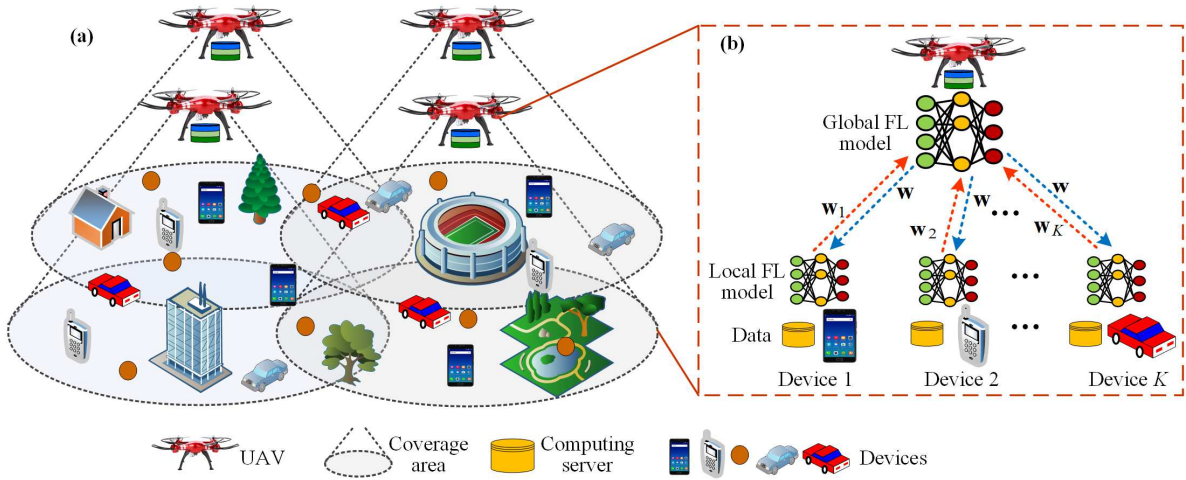}
\caption{{Federated learning-based UAV-enabled wireless networks.} } \label{fig:Schematic}
\end{figure}

In the FL instance, each device has its own personal dataset and it is
willing to upload a part of inflammation (i.e., AI model parameters) to its associated UAV server
in a privacy-preserving manner. In addition, devices involved
in a common computing task (e.g., training a classification model)
are more likely to work with others together to finish the task
collaboratively, by adopting AI techniques. These
devices will update their own local AI model
parameters to associated UAV servers for global model aggregation.
Each device processes the local model training based on its local
raw dataset without sharing the raw data with other devices, to
protect the privacy of data providers. After completing local
model training, each device will send its local model parameters
to its associated UAV server by the uplink communication channel,
and the corresponding UAV server aggregates the local parameters
from the selected devices before broadcasting the aggregated
global parameters to the devices by the downlink communication
channel.

For the FL aggregation, communication and computation
resource optimization (related to both the computation capability
and communication capability) is necessary to realize the
implementation of model aggregation, such as local computation,
communication [5], [6], and global computation at both the devices and the
UAV servers. The computation capacity of each device or each UAV
server is captured by the CPU capability, learning time, and
learning accuracy. The communication capacity can be
characterized in the forms of the available transmission rate and
transmission latency. It is worth noting that both the computation
and communication capabilities may vary at different time slots
due to different computing tasks and mobility of UAVs or devices.
Hence, this paper also focuses on joint device selection, UAVs
placement design, and resource management to
enhance the leaning efficiency and accuracy of FL.

\subsection{FL Model}

This subsection briefly investigates the basics of FL in
multi-UAV-enabled networks. Hereinafter, the considered AI model that is trained on each device's training dataset
is called  local FL model, while the FL model that is built by
each UAV server employing local model parameter inputs from its
selected devices is called global FL model.

Let ${{\bf{w}}_n}$ denote the model parameters which is related to
the global model of the $n$-th UAV server, let ${{\bf{w}}_k}$
denote the local model parameters of the $k$-th device, and let
${\mathcal{D}_k}$ denote the set of training dataset used at the
$k$-th device. Accordingly, if the $k$-th device is associated with
the $n$-th UAV server, we introduce the loss function $f\left(
{{{\bf{w}}_n};{{\bf{s}}_{k,i}},{z_{k,i}}} \right)$ to quantify the
FL performance error over the input data sample vector
${{\bf{s}}_{k,i}}$ on the learning model ${\bf{w}}$ and the
desired output scalar ${z_{k,i}}$ for each input sample $i$ at the
$k$-th device. For the $k$-th device, the sum loss function on its
training dataset ${\mathcal{D}_k}$ can be expressed as [5], [6]
\begin{equation}
\begin{split}
{F_k}\left( {{{\bf{w}}_n}} \right) = \frac{1}{{\left|
{{{\mathcal{D}_k}}} \right|}}\sum\limits_{i \in {{\mathcal{D}_k}}}
{f\left( {{{\bf{w}}_n};{{\bf{s}}_{k,i}},{z_{k,i}}} \right)}
,\;\;\forall k \in \mathcal{K},
\end{split}
\end{equation}
where  $\left| {{\mathcal{D}_k}} \right|$ is the cardinality of
set  ${\mathcal{D}_k}$. Accordingly, at the $n$-th UAV server, the
average global loss function with the distributed local datasets
of all selected devices is defined as [5], [6]
\begin{equation}
\begin{split}
F\left( {{{\bf{w}}_n}} \right) \buildrel \Delta \over =
\sum\limits_{k \in {\mathcal{K}_n}} {\frac{{\left|
{{\mathcal{D}_k}} \right|{F_k}({{\bf{w}}_n})}}{{\left|
{{\mathcal{D}_n}} \right|}}} = \frac{1}{{\left| {{\mathcal{D}_n}}
\right|}}\sum\limits_{k \in {\mathcal{K}_n}} {\sum\limits_{i \in
{\mathcal{D}_k}} {f\left(
{{{\bf{w}}_n};{{\bf{s}}_{k,i}},{z_{k,i}}} \right)} } ,
\end{split}
\end{equation}
where $\left| {{\mathcal{D}_n}} \right| = \sum\nolimits_{k \in
{{\mathcal{K}}_n}} {\left| {{\mathcal{D}_k}} \right|} $ is the sum data
samples from all selected devices at the $n$-th UAV-enabled cell, and
${\mathcal{K}_n}$ is the set of the devices associated with the
$n$-th UAV server with ${K_n} = |{\mathcal{K}_n}|$ being the
number of the selected devices. The objective of the FL task is to
search for the optimal model parameters at the $n$-th UAV server that
minimizes the global loss function as [5], [6]
\begin{equation}
\begin{split}
{\bf{w}}_n^ *  = \arg \min F\left( {{{\bf{w}}_n}}
\right),\;\forall n \in \mathcal{N}.
\end{split}
\end{equation}

Note that $F\left( {{{\bf{w}}_n}} \right)$ cannot be directly
computed by employing the raw datasets from selected devices, for the sake of protecting the information privacy and security of each
device. Hence, the problem in (3) can be addressed in a
distributed manner. We pay attention to the FL model-averaging
implementation in the subsequent exposition while the same
principle is generally adopted to realize alternative
implementation according to the gradient-averaging method [5], [6].

\subsection{Communication Model}

Different from the propagation of ground communications, the
air-to-ground or ground-to-air channel mainly depends on the
propagation environments, transmission distance, and elevation
angle [1]-[3]. Similar to existing works [23]-[28], the UAV-to-device or
device-to-UAV communication channel link is modeled by a
probabilistic path loss model, where both the LoS and non-LoS
(NLoS) links are taken into account. The LoS and NLoS path loss in
dB from the $n$-th UAV server to the $k$-th device can be given by
\begin{equation}
\begin{split}
l_{n,k}^{_{{\rm{LoS}}}} = 20\log \left( {{{4\pi {f_c}{d_{n,k}}}
\mathord{\left/
 {\vphantom {{4\pi {f_c}{d_{n,k}}} c}} \right.
 \kern-\nulldelimiterspace} c}} \right) + {\eta ^{{\rm{LoS}}}},
\end{split}
\end{equation}
\begin{equation}
\begin{split}
l_{n,k}^{_{{\rm{NLoS}}}} = 20\log \left( {{{4\pi {f_c}{d_{n,k}}}
\mathord{\left/
 {\vphantom {{4\pi {f_c}{d_{n,k}}} c}} \right.
 \kern-\nulldelimiterspace} c}} \right) + {\eta ^{{\rm{NLoS}}}},
\end{split}
\end{equation}
respectively, where ${f_c}$ and $c$  denote the carrier frequency
and the speed of light, respectively. ${d_{n,k}}$ is the
transmission distance from the $n$-th UAV server to the $k$-th device,
given by  ${d_{n,k}} = {\left[ {{{({x_n} - {x_k})}^2} + {{({y_n} -
{y_k})}^2} + h_n^2} \right]^{{1 \mathord{\left/
 {\vphantom {1 2}} \right.
 \kern-\nulldelimiterspace} 2}}}$, where ${\Theta _n} = ({x_n},{y_n},{h_n})$ denotes the location of the $n$-th UAV
server,  ${\Theta _k} = ({x_k},{y_k})$ is the location of the
$k$-th device, and $h$ is the height of UAV servers in the sky, and we assume that all UAVs have the same height. ${\eta ^{{\rm{LoS}}}}$ and  ${\eta
^{{\rm{NLoS}}}}$ are the mean additional losses for the LoS and
NLoS links due to the free space propagation loss, respectively,
as defined in [29]. In the communication model, the
probability of LoS connection between the $n$-th UAV server and the
$k$-th device is expressed as
\begin{equation}
\begin{split}
{\rm{P}}_{n,k}^{_{{\rm{LoS}}}} = \frac{1}{{1 + {\xi _1}\exp \left(
{ - {\xi _2}({\theta _{n,k}} - {\xi _1})} \right)}},
\end{split}
\end{equation}
where ${\xi _1}$ and ${\xi _2}$ are constant values which depend
on the carrier frequency and UAV network environment, and ${\theta
_{n,k}}$ is the elevation angle from UAV $n$ to device $k$ (in
degree): ${\theta _{n,k}} = {\textstyle{{180} \over {\rm{\pi }}}}
\cdot {\sin ^{ - 1}}({\textstyle{{{h_n}} \over {{d_{n,k}}}}})$.
Furthermore, the probability of NLoS is
${\rm{P}}_{n,k}^{_{{\rm{NLoS}}}} = 1 -
{\rm{P}}_{n,k}^{_{{\rm{LoS}}}}$. In this context, the
probabilistic path loss between the $n$-th UAV and the $k$-th device
is given by
\begin{equation}
\begin{split}
{l_{n,k}} = {\rm{P}}_{n,k}^{_{{\rm{LoS}}}} \cdot
l_{n,k}^{_{{\rm{LoS}}}} + {\rm{P}}_{n,k}^{_{{\rm{NLoS}}}} \cdot
l_{n,k}^{_{{\rm{NLoS}}}}.
\end{split}
\end{equation}

We assume that orthogonal frequency-division multiple access
(OFDMA) technique is used for uplink channel access, where each
UAV-enabled cell has $M$ orthogonal uplink subchannels and these
subchannels are reused across all cells. Note that the total uplink bandwidth is
divided into $M$ orthogonal subchannels, denoted by $\mathcal{M} =
\{ 1,...,M\} $. In this case, each UAV server will suffer
inter-cell interference (ICI) from other nearby devices associated
by other cells on the same spectrum band. Hence, according to the path loss model, when the
$k$-th device is associated with the $n$-th UAV server, the
received signal-to-interference plus-noise-ratio (SINR) over the
allocated subchannel $m$ at the $n$-th UAV server in the uplink is
characterized as
\begin{equation}
\begin{split}
SINR_{n,k,m}^{\rm{U}} = \frac{{P_{k,m}^{\rm{U}}{{10}^{ -
{l_{n,k}}/10}}}}{{\sum\limits_{k' \in \mathcal{K},k' \ne k}
{P_{k',m}^{\rm{U}}{{10}^{ - {l_{n,k}}/10}}}  + {\sigma ^2}}},
\end{split}
\end{equation}
where $P_{k,m}^{\rm{U}}$ is the transmit power of the $k$-th
device allocated on the $m$-th subchannel, and   ${\sigma ^2}$ denotes
the power of the Gaussian noise. Furthermore, $\sum\limits_{k' \in
\mathcal{K},k' \ne k} {P_{k',m}^{\rm{U}}{{10}^{ - {l_{n,k}}/10}}}
$ is the ICI received at UAV server $n$ over the $m$-th subchannel
which is generated from nearby devices associated by other cells.
Accordingly, the achievable uplink data rates (in bit/s) for the
$k$-th device over the allocated subchannels can be expressed as
\begin{equation}
\begin{split}
R_k^{\rm{U}} = {B_{{\rm{sub}}}}\sum\limits_{m = 1}^M {\left(
{{\chi _{n,k,m}}{{\log }_2}(1 + SINR_{n,k,m}^{\rm{U}})} \right)},
\end{split}
\end{equation}
where ${B_{{\rm{sub}}}} = {B^{\rm{U}}}/M$ is the bandwidth of each
uplink subchannel with ${B^{\rm{U}}}$ being the total bandwidth in
uplink, and  ${\chi _{n,k,m}}$ is the uplink subchannel allocation
indicator,   ${\chi _{n,k,m}} \in \{ 0,1\} $; ${\chi _{n,k,m}} =
1$  shows that the $k$-th device is associated with the $n$-th UAV
server on the $m$-th subchannel ; otherwise,  ${\chi _{n,k,m}} =
0$.

For the downlink channel, we assume that each UAV server occupies
a given downlink channel to broadcast the global model
parameters to its associated devices. As the deployment of multiple
UAVs in the sky, devices located in the overlapped areas will also
suffer ICI from other nearby UAV servers on the same spectrum
band. Then, when the $k$-th device is associated with the $n$-th UAV
server, its received SINR in the downlink can be expressed as
\begin{equation}
\begin{split}
SINR_{n,k}^{\rm{D}} = \frac{{P_n^{\rm{D}}{{10}^{ -
{l_{n,k}}/10}}}}{{\sum\limits_{n' \in \mathcal{N}, n' \ne n}
{P_{n'}^{\rm{D}}{{10}^{ - {l_{n,k}}/10}}}  + {\sigma ^2}}}.
\end{split}
\end{equation}
where $P_n^{\rm{D}}$ is the transmit power of the $n$-th UAV
server. Accordingly, the achievable  downlink data rate at the
$k$-th device is given by
\begin{equation}
\begin{split}
R_k^{\rm{D}} = {B^{\rm{D}}}{\log _2}(1 + SINR_{n,k}^{\rm{D}}),
\end{split}
\end{equation}
where ${B^{\rm{D}}}$ is the bandwidth in the downlink.

\subsection{FL Model Update Latency Analysis}

\textbf{Local/Global Model Update Latency:} Define ${C_k}$ and
${C_n}$ as the number of CPU cycles used for  training model on
one sample data at the $k$-th device and $n$-th UAV server,
respectively. Let ${f_k}$ and ${f_n}$ denote the computation
capability (CPU cycles/s) of device $k$ and UAV server $n$,
respectively, where ${f_k} \in (f_k^{\min },f_k^{\max })$ with
$f_k^{\min }$ and $f_k^{\max }$ being the minimum and maximum CPU
computation capabilities of device $k$, respectively. Accordingly,
the local model computation latency of device $k$ and the global
model aggregation latency of UAV server $n$ at the $t$-th time
slot are respectively given by
\begin{equation}
\begin{split}
T_k^{{\rm{Loc}},t} = \left| {{\mathcal{D}_k}}
\right|{C_k}/{f_k},\;\forall k,
\end{split}
\end{equation}
\begin{equation}
\begin{split}
T_n^{{\rm{Glo}},t} = \left| {{\mathcal{D}_n}}
\right|{C_n}/{f_n},\;\forall n.
\end{split}
\end{equation}

\textbf{Global FL Model Broadcast Latency:} Let ${L_n}$  denote
the number of bits required for each UAV server $n$ to broadcast the global
 model parameters to the associated devices. For the $n$-th UAV server,
the global model parameters broadcast latency is expressed as
\begin{equation}
\begin{split}
T_n^{{\rm{D}},t} = {L_n}/R_n^{{\rm{D}},t},\;\forall n.
\end{split}
\end{equation}

\textbf{Local FL Model Upload Latency:} Let ${L_k}$  denote the
number of bits needed for each device $k$ to upload its local
model parameters to its associated UAV server. For the $k$-th device, the
local model parameters upload latency can be given by
\begin{equation}
\begin{split}
T_k^{{\rm{U}},t} = {L_k}/R_k^{{\rm{U}},t},\;\forall k.
\end{split}
\end{equation}

The one round time for scheduling the $k$-th device is comprised
of local model update latency, uplink local model upload latency,
global model aggregation latency, and downlink global model
broadcast latency. Therefore, the total time cost for scheduling
the FL-model of the $k$-th device at one round can be given by
\begin{equation}
\begin{split}
T_k^t = T_k^{{\rm{Loc}},t} + T_n^{{\rm{Glo}},t}
+ T_k^{{\rm{D}},t}+ T_k^{{\rm{U}},t},\;\;\forall k.
\end{split}
\end{equation}

\subsection{Problem Formulation}

In UAV-enabled networks, the various computation capacities of
devices and the time-varying communication channel conditions play
an important role on the implementation of FL global
aggregation. It is desirable to select the devices with high
computation capability, communication capability, and accurate learned
models. In addition, it is necessary to schedule UAVs' locations
to provide the best channel gains for communication services, and
manage the radio and power resources to improve the communication
data rate for AI model parameters upload and broadcast. Thus, we
aim to select a subset of devices, design UAVs' locations,
manage subchannel and transmit power resources to minimize the FL
model execution time and the learning accuracy loss. In the $n$-th
UAV-enabled cell, we define the execution time cost  as
\begin{equation}
\begin{split}
c_n^{{\rm{Time}}}(t) = \frac{1}{{{K_n}}}\sum\limits_{k =
1}^{{K_n}} {T_k^t} ,\;\;\forall n,
\end{split}
\end{equation}
where ${K_n}$ is the number of devices selected by the $n$-th UAV
server for the federated model aggregation. In addition, the learning accuracy loss can be defined as
\begin{equation}
\begin{split}
c_n^{{\rm{Loss}}}(t) = \frac{1}{{\left| {{\mathcal{D}_n}}
\right|}}\sum\limits_{k \in {\mathcal{K}_n}} {\sum\limits_{i \in
{\mathcal{D}_k}} {f\left(
{{{\bf{w}}_n};{{\bf{s}}_{k,i}},{z_{k,i}}} \right)} } ,\;\;\forall
n.
\end{split}
\end{equation}
In the network, the learning accuracy loss is measured at the end
of each time slot.

Given the aforementioned system model, our objective is to
minimize the weighted sum of one-round FL model execution time and
learning accuracy loss. Targeting  at learning
acceleration and efficiency, it is desirable to select a subset of
devices with high computation capability, place the UAVs'
locations with best channel quality, as well as manage both
subchannel and power resources. Hence, the optimization problem
can be formulated as follows
\begin{equation}
\begin{split}
\begin{array}{l}
\mathop {\min }\limits_{{\Theta _{n,}}{\bm{\rho }},{\bm{\chi }},P_n^{\rm{D}}} \left( {\lambda c_n^{{\rm{Time}}}(t) + (1 - \lambda )c_n^{{\rm{Loss}}}(t)} \right),\;\;\forall n\\
s.t.\;\;{\rm{a}})\;{\rho _{n,k}} \in \{ 0,1\} ,\;{\chi _{n,k,m}} \in \{ 0,1\} ,\forall k,m,\\
\;\;\;\;\;\;\;{\rm{b}})\sum\limits_{n \in \mathcal{N}} {{\rho _{n,k}} \le 1,\forall k} ,\\
\;\;\;\;\;\;\;{\rm{c}})\sum\limits_{k \in \mathcal{K}} {\sum\limits_{m \in M} {{\chi _{n,k,m}} \le M,} } \\
\;\;\;\;\;\;\;{\rm{d}})\;0 \le P_n^{\rm{D}} \le P_n^{{\rm{max}}},\\
\;\;\;\;\;\;\;{\rm{e}})\;0 \le {f_k} \le f_k^{\max },\;\forall k,
\end{array}
\end{split}
\end{equation}
where  ${\bm{\rho }} = {[{\rho _{n,1}}, \ldots ,{\rho _{n,K}}]^T}$
and  ${\bm{\chi }} = [{\chi _{n,1,1}}, \ldots ,{\chi
_{n,1,M}},{\chi _{n,2,1}}, \ldots ,{\chi _{n,2,M}}, \ldots ,{\chi
_{n,K,M}}]$,  $P_n^{{\rm{max}}}$ is the maximum transmit power of
UAV server $n$, and  $\lambda  \in [0,1]$ is a constant weight
parameter which is used to balance the one-round execution time
$c_n^{{\rm{Time}}}$ and the learning accuracy loss
$c_n^{{\rm{Loss}}}$. Constraint (19b) indicates that each device
can only associate one UAV server to preform model aggregation.
Constraint (19c) is used to ensure the maximum number of available
subchannels of each UAV-enabled cell. Constraint (19d) ensures the
maximum transmit power of UAV server $n$ in the downlink. Constraint
(19d) represents the computation capacity range of devices. We
would like to mention that for the resource management issue, we
focus our study on the uplink subchannel allocation since the
global model parameters at each UAV server are broadcast by a
given downlink band. In addition, we also investigate the power allocation at each
UAV server by assuming that the transmit power of each device is
given.

\section{Asynchronous Federated Reinforcement Learning Solution}

The optimization problem formulated in (19) is challenging to tackle
as it is a non-convex combination and NP-hard problem. In
addition, the time-varying channel condition and different
computation capacities of devices result in dynamic open and
uncertain characteristics, which increases the difficulty of
addressing the optimization problem. Model-free RL is one of the
dynamic programming technique which is capable of tackling the
decision-making problem by learning an optimized policy in dynamic
environments [30]. Thus, RL is introduced to implement the
self-scheduling based FL aggregation process in multi-UAV-enable
networks.

Moreover, traditional FL models mostly use a synchronous
learning framework to update the model parameters  between the
UAV servers and client devices, which inherently have several key challenges.
Firstly, due to the mobility and different computation capacities of
mobile devices, it is difficult to maintain continuous
synchronized communication between UAV servers and client devices.
Secondly, each UAV server needs to wait for all selected devices to
finish their local results before aggregating the  model
parameters, which consequently increases the global learning delay
with low round efficiency. Furthermore, a part of the action
(i.e., transmission power and horizontal locations of UAVs) in our
joint placement design and resource management optimization
problem have continuous spaces. Therefore, we propose an asynchronous advantage actor-critic-based  asynchronous
federated learning algorithm (called A3C-AFL) to address the aforementioned
challenges, and the corresponding framework is provided with the
following extensive details in this section.

\begin{figure}
\centering
\includegraphics[width=1\columnwidth]{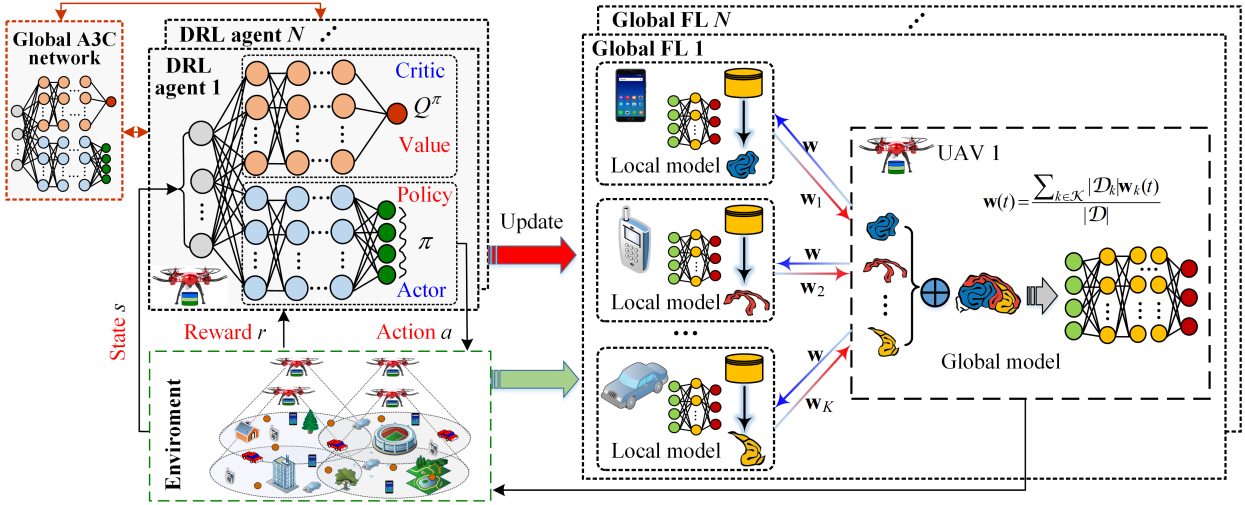}
\caption{{The framework of A3C-AFL in UAV-enabled networks.} } \label{fig:Schematic}
\end{figure}

The proposed A3C-AFL framework consists of three phases: device
selection, UAVs placement as well as resource management; local
training; and global aggregation, as illustrated in Fig. 2. The
strategy of the device selection, UAV placement design, and
resource management can be achieved by using the A3C algorithm. Then, the
selected devices in each UAV-enabled cell perform local training
and periodically upload their local  model parameters to their
corresponding associated UAV server for global aggregation. Finally, each UAV server broadcasts the updated global model parameters to the
associated devices.

\begin{itemize}
\item \textbf{Device Selection, UAVs Placement, and Resource Management:}
To improve the leaning efficiency and accuracy, each UAV server
selects the devices with high computation and communication
capacities to perform federated model parameters update. Then, the joint
UAVs placement design and resource management are implemented to
minimize the execution time. This phase is implemented by adopting the A3C-based algorithm [31],
which will be elaborated in Section III.B.

\item \textbf{Local Training:} At the $t$-th global communication round,
after receiving the global model parameters ${{\bf{w}}_n}(t - 1)$
from the associated UAV server, each selected device $k$ trains
its local model parameters ${{\bf{w}}_k}(t)$ based on its dataset
${\mathcal{D}_k}$  by calculating the local stochastic gradient
descent $\nabla {F_k}\left( {{{\bf{w}}_n}(t - 1)} \right)$, i.e.,
\begin{equation}
\begin{split}
{{\bf{w}}_k}(t) = {{\bf{w}}_n}(t - 1) - \eta \nabla {F_k}\left(
{{{\bf{w}}_n}(t - 1)} \right),
\end{split}
\end{equation}
where $\eta  > 0$ denotes the learning step size and $\nabla $ is
the gradient operator. After updating the parameters
${{\bf{w}}_k}(t)$, each device $k$ uploads its trained local model parameters ${{\bf{w}}_k}(t)$ to its associated UAV server for further
FL model aggregation.

\item  \textbf{Global Aggregation:} Each UAV server $n$ (aggregator)
retrieves the uploaded local FL model parameters from its selected
devices and implements the global model aggregation through
averaging and updating the global model parameters
${{\bf{w}}_n}(t)$ as follows
\begin{equation}
\begin{split}
{{\bf{w}}_n}(t) = \frac{{\sum\nolimits_{k \in {K_n}}
{|{\mathcal{D}_k}|{{\bf{w}}_k}(t)} }}{{\left| {{\mathcal{D}_n}} \right|}}.
\end{split}
\end{equation}

\end{itemize}

The mode parameters update process involves the iteration between
(20) and (21) until the FL model coverages.

\subsection{Modeling of Reinforcement Learning Environment}

We formulate the combinatorial optimization problem (19) as a
Markov decision process (MDP), i.e.,  $\left\langle
{\mathcal{S},\mathcal{A},\mathcal{P},r} \right\rangle $, where
$\mathcal{A}$, $\mathcal{S}$, $\mathcal{P}$, and $r$ are network state
space, action space, state transition probability, and reward
function, respectively. In multi-UAV-enabled networks, UAV servers
are considered as multiple agents to observe the environment and
try to maximize the expected sum reward. Then, we use RL to tackle
the device selection, UAVs placement, and resource management
problem (19) in the federated learning scenario. All of the multiple
agents iteratively update their policies based on their observations
by interacting with the network environment. According to the
aforementioned MDP, i.e.,  $\left\langle
{\mathcal{S},\mathcal{A},\mathcal{P},r} \right\rangle $, the
corresponding elements in MDP are described as below:

\textbf{State Space:} At each time slot $t$, the network state
${s_n}(t) \in \mathcal{S}$ of each UAV server $n$ for characterizing the
environment is comprised of the following several parts:
horizontal location of the $n$-th UAV server ${\Theta _n}(t) =
\left( {{x_n}(t),{y_n}(t)} \right)$; device selection indicators at
the $n$-th UAV-enabled cell, ${\left\{ {{\rho _{n,k}}(t
- 1)} \right\}_{k \in {\mathcal{K}_n}}}$; the subchannel
allocation state ${\left\{ {{\chi _{n,k,m}}(t - 1)} \right\}_{k
\in {\mathcal{K}_n}}}$; locations of selected devices, ${\left\{
{{\Theta _k}(t) = \left( {{x_k}(t),{y_k}(t)} \right)} \right\}_{k
\in {\mathcal{K}_n}}}$; the remaining payload needs to be
transmitted, ${\left\{ {{L_k}(t - 1)} \right\}_{k \in {K_n}}}$ and
${L_n}(t - 1)$. According to the above definitions, in the multi-agent
RL model, the network state of the $n$-th UAV server at time slot
$t$ can be expressed by
\begin{equation}
\begin{split}
{s_n}(t) = \left\{ {{\Theta _n}(t),{L_n}(t - 1),{{\left\{ {{\Theta
_k}(t),{\rho _{n,k}}(t - 1),{\chi _{n,k,m}}(t - 1),{L_k}(t)}
\right\}}_{k \in {\mathcal{K}_n}}}} \right\},
\end{split}
\end{equation}
and the network state of all UAV servers is given by  ${s_t} =
\left\{ {{{\{ {s_n}(t)\} }_{n \in \mathcal{N}}}} \right\}$.

\textbf{Action Space:} At the $t$-th time slot, each agent $n$
(i.e., UAV server $n$) selects its corresponding action ${a_n}(t)
\in \mathcal{A}$ according to the observed state ${s_n}(t)$, where
${a_n}(t)$ consists of the horizontal position ${\Theta _n}(t) =
\left( {{x_n}(t),{y_n}(t)} \right)$, the device selection
indicators  ${\left\{ {{\rho _{n,k}}(t)} \right\}_{k \in
{\mathcal{K}_n}}}$, the subchannel allocation indicators ${\left\{
{{\chi _{n,k,m}}(t)} \right\}_{k \in {\mathcal{K}_n}}}$, and the
transmit power allocation level $P_n^{\rm{D}}(t)$, that is
\begin{equation}
\begin{split}
{a_n}(t) = \left\{ {{\Theta _n}(t),P_n^{\rm{D}}(t),{{\left\{
{{\rho _{n,k}}(t),{\chi _{n,k,m}}(t)} \right\}}_{k \in
{\mathcal{K}_n}}}} \right\},
\end{split}
\end{equation}
and the action of all UAV servers is given by ${a_t} = \left\{
{{{\{ {a_n}(t)\} }_{n \in \mathcal{N}}}} \right\}$. At the end of
each time slot $t$, each UAV server moves to an updated horizontal
position, updates the device association indicators, allocates subchannel and power resources to devices, respectively.

\textbf{State Transition Function:} Let $\mathcal{P}({s_{t +
1}}|{s_t},{a_t})$ denote the transition probability of one UAV
server entering a new state  ${s_{t + 1}}$, after executing an
action ${a_t}$ at the current state  ${s_t}$.

\textbf{Policy:} Let $\pi$ denote the policy function, which is a mapping from perceived states with the probability distribution over actions that the agent can select in those states, $\pi(a|s)=\mathcal{P}(a|s)$.

\textbf{Reward Function:} In the context of minimizing the
one round FL model execution time and learning accuracy loss,
the reward function ${r_t}$  is designed to evaluate
the quality of a learning policy under the current state-action
pair $({s_t},{a_t})$. In this paper, we design a reward function that can capture
the federated execution time and learning accuracy loss. According to the objective function in (19), the presented
reward function of the $n$-th agent (i.e., UAV server $n$) at
one time step is expressed as
\begin{equation}
\begin{split}
{r_n}({s_t},{a_t}) =  - \left( {\lambda c_n^{{\rm{Time}}}(t) + (1
- \lambda )c_n^{{\rm{Loss}}}(t)} \right).
\end{split}
\end{equation}

The objective of the UAV-enabled network is to minimize the
execution time and learning accuracy loss in FL. For the MDP model,
the objective is to search for an action $a$ which is capable of
maximizing the cumulative reward (minimize the total cumulative
execution time and learning  accuracy loss), given by
\begin{equation}
\begin{split}
{U_t} = \sum\limits_{\tau  = 0}^\infty  {{\gamma _\tau }r({s_{t +
\tau }},{a_{t + \tau }})},
\end{split}
\end{equation}
where $\gamma$ denotes the discount factor.

\subsection{Multi-Agent A3C Algorithm}
In multi-UAV-enabled networks, the spaces
of both state and action are large, and hence combining deep
learning and RL, i.e., DRL algorithms (e.g., deep Q-learning and deep
deterministic policy gradient), are effective for handling the
large-scale decisions making problems [30]. These algorithms generally
adopt experience replay to improve learning efficiency. However,
experience replay requires enough memory space and computation
resources to guarantee the learning accuracy, and it only uses the
data generated through old policy to update the learning process [31]. The negative
issues of the aforementioned DRL algorithms motivate  us to search for a
better algorithm, A3C (asynchronous advantage actor-critic).

Different from a classical AC algorithm with only one
learning agent, A3C is capable of enabling asynchronous multiple
agents (i.e., UAV servers) to parallelly interact with their
environments and achieve different exploration policies. In
A3C, the actor network generally adopts policy gradient schemes to
select actions under a given parameterized policy $\pi
(a|s;{{\bm{\theta }}_{\rm{a}}}) = \mathcal{P}(a|s,{{\bm{\theta
}}_{\rm{a}}})$ with a set of actor parameters ${{\bm{\theta
}}_{\rm{a}}}$, and then updates the parameters by the
gradient-descent methods. The critic network uses an estimator of
the state value function ${V^\pi }({s_t};{{\bm{\theta
}}_{\rm{c}}})$ to qualify the expected return under a certain
state $s$ with a set of critic parameters  ${{\bm{\theta
}}_{\rm{c}}}$.

Each UAV server acts as one agent to evaluate and optimize its
policy based on the value function, which is defined as the
expected long-term cumulative reward achieved over the entire
learning process. Here, we defined two value functions, called
state value function and state-action value function, where the
former one is the expected return under a given policy $\pi $
while the latter one is the expected return under a given policy
$\pi $ after executing action $a$ in state $s$. These two
functions are respectively expressed as
\begin{equation}
\begin{split}
{V^\pi }({s_t};{{\bm{\theta }}_{\rm{c}}}) = \mathbb{E}\left\{
{{U_t}|{s_t} = s,\pi } \right\},
\end{split}
\end{equation}
\begin{equation}
\begin{split}
{Q^\pi }({s_t},{a_t}) = \mathbb{E}\left\{ {{U_t}|{s_t} = s,{a_t} =
a,\pi } \right\},
\end{split}
\end{equation}
where $\mathbb{E}\{  \cdot \} $ is the expectation.

A3C adopts the multi-step reward to update the parameters of the
policy in the actor network and the value function in the critic
network [31]. Here, the $i$-step reward is defined as [31]
\begin{equation}
\begin{split}
{U_t} = \sum\limits_{i = 0}^{\tau  - 1} {{\gamma _i}r({s_{t +
i}},{a_{t + i}}) + {\gamma _\tau }{V^\pi }({s_{t + \tau
}};{{\bm{\theta }}_{\rm{c}}})} .
\end{split}
\end{equation}

Similar to the AC framework, A3C also adopts policy gradient
schemes to perform parameters update which may cause high variance
in the critic network. In order to address this issue, an
advantage function $A(s,a) = {Q^\pi }(s,a) - {V^\pi
}({s_t};{{\bm{\theta }}_{\rm{c}}})$ is employed to replace ${Q^\pi
}(s,a)$ in the policy gradient process. Since ${Q^\pi }(s,a)$
cannot be determined in A3C [31], we use ${U_t} - {V^\pi
}({s_t};{{\bm{\theta }}_{\rm{c}}})$ as an estimator for the
advantage function $A(s,a) = {Q^\pi }(s,a) - {V^\pi
}({s_t};{{\bm{\theta }}_{\rm{c}}})$. As a result, the advantage
function is given by
\begin{equation}
\begin{split}
\begin{array}{l}
A({s_t},{a_t}) = {U_t} - {V^\pi }({s_t};{{\bm{\theta }}_{\rm{c}}})\\
 = \sum\limits_{i = 0}^{\tau  - 1} {{\gamma _i}r({s_{t + i}},{a_{t + i}}) + {\gamma _\tau }{V^\pi }({s_{t + \tau }};{{\bm{\theta }}_{\rm{c}}}) - {V^\pi }({s_t};{{\bm{\theta }}_{\rm{c}}})}.
\end{array}
\end{split}
\end{equation}

In the A3C framework, two loss functions are associated with the
two deep neural network outputs of the actor network and the
critic network. According to the advantage function (29), the actor
loss function [31] under a given policy $\pi$ is defined as
\begin{equation}
\begin{split}
{f_\pi }({{\bm{\theta }}_{\rm{a}}}) = \log \pi
({a_t}|{s_t};{{\bm{\theta }}_{\rm{a}}})\left( {{U_t} - {V^\pi
}({s_t};{{\bm{\theta }}_{\rm{c}}})} \right) + \vartheta G\left(
{\pi ({s_t};{{\bm{\theta }}_{\rm{a}}})} \right),
\end{split}
\end{equation}
where $\vartheta $ is a hyperparameter that controls the strength
of the entropy regularization term, such as the exploration and
exploitation management during the training process, and $G\left(
{\pi ({s_t};{{\bm{\theta }}_{\rm{a}}})} \right)$ is the entropy
which is employed to favor exploration in training. From (30),  the
accumulated gradient of the actor loss function ${f_\pi
}({{\bm{\theta }}_{\rm{a}}})$ is expressed as
\begin{equation}
\begin{split}
d{{\bm{\theta }}_{\rm{a}}} = d{{\bm{\theta }}_{\rm{a}}} + {\nabla
_{{{{\bm{\theta '}}}_{\rm{a}}}}}\log \pi ({a_t}|{s_t};{{\bm{\theta
'}}_{\rm{a}}})\left( {{U_t} - {V^\pi }({s_t};{{\bm{\theta
}}_{\rm{c}}})} \right) + \vartheta {\nabla _{{{{\bm{\theta
'}}}_{\rm{a}}}}}G\left( {\pi ({s_t};{{{\bm{\theta '}}}_{\rm{a}}})}
\right),
\end{split}
\end{equation}
where ${{\bm{\theta '}}_{\rm{a}}}$ is the thread-specific actor
network parameters.

In addition, the critic loss function of the estimated value
function is defined as
\begin{equation}
\begin{split}
f({{\bm{\theta }}_{\rm{c}}}) = {\mathop{\rm}\nolimits} {\left(
{{U_t} - {V^\pi }({s_t};{{\bm{\theta }}_{\rm{c}}})} \right)^2},
\end{split}
\end{equation}
and the accumulated gradient of the critic loss function
$f({{\bm{\theta }}_{\rm{c}}})$ in the critic network is calculated
by
\begin{equation}
\begin{split}
d{{\bm{\theta }}_{\rm{c}}} = d{{\bm{\theta }}_{\rm{c}}} +
\frac{{\partial {{\left( {{U_t} - {V^\pi }({s_t};{{\bm{\theta
}}_{\rm{c}}})} \right)}^2}}}{{\partial {{{\bm{\theta
'}}}_{\rm{c}}}}},
\end{split}
\end{equation}
where ${{\bm{\theta '}}_{\rm{c}}}$ is the  thread-specific critic
network parameters.

After updating the accumulated gradients shown in (31) and (33), we
adopt the standard non-centered RMSProp algorithm [32] to perform
training for the loss function minimization in our presented A3C
framework. The estimated gradient with the RMSProp algorithm of
the actor or critic network is expressed as
\begin{equation}
\begin{split}
g = \alpha g + (1 - \alpha ){(\Delta {\bm{\theta }})^2},
\end{split}
\end{equation}
where $\alpha $ denotes the momentum, and $\Delta {\bm{\theta }}$
is the accumulated gradient ($d{{\bm{\theta }}_{\rm{a}}}$ or
$d{{\bm{\theta }}_{\rm{c}}}$) of the actor loss function or the
critic loss function. Note that the estimated gradient $g$ can be
either shared or separated across agent threads, but the shared
mode tends to be more robust [31], [32]. The estimated gradient
$g$ is used to update the parameters of both the actor and critic
networks as
\begin{equation}
\begin{split}
{\bm{\theta }} \leftarrow {\bm{\theta }} - \beta \frac{{\Delta
{\bm{\theta }}}}{{\sqrt {g + \varepsilon } }},
\end{split}
\end{equation}
where $\beta $ is the learning rate, and $\varepsilon $ is a small
positive step.

\subsection{Training and Execution of A3C-AFL}

Like most of machine learning algorithms, there are two stages in
the proposed A3C-AFL algorithm, i.e., the training procedure and
the execution procedure. Both the training and execution datasets
are generated from their interaction of a federated learning
environment conducted by the multi-UAV-enabled wireless network.

\linespread{1.1}{
\begin{algorithm}[t]
\begin{normalsize}
\caption{\normalsize A3C-based Device Selection, UAVs Placement,
and Resource Management}
1:$~$ \textbf{Initialization:} Initialize the global parameters
${{\bm{\theta }}_{\rm{a}}}$ and ${{\bm{\theta }}_{\rm{c}}}$
of the actor and critic networks in the global network;\\
$~~~~$ Initialize the thread-specific parameters ${{\bm{\theta
'}}_{\rm{a}}}$  and ${{\bm{\theta '}}_{\rm{c}}}$ in the local
networks; \\
$~~~$ Initialize the global shared counter  $T = 0$ and thread counter
$t = 1$ with
the maximum counters  $T_{\max }^{{\rm{A3C}}}$ and $t_{\max }^{{\rm{A3C}}}$, respectively;\\
2:$~$ Set  $\alpha ,\beta ,\gamma ,\varepsilon $, $T_{\max
}^{{\rm{A3C}}}$, and $t_{\max }^{{\rm{A3C}}}$, respectively;\\
3:$~$ \textbf{while} $T < T_{\max }^{{\rm{A3C}}}$ \textbf{do}\\
4:$~~~$ \textbf{for} each learning agent \textbf{do}\\
5:$~~~$ Reset two accumulated gradients:  $d{{\bm{\theta
}}_{\rm{a}}} \leftarrow 0$ and  $d{{\bm{\theta }}_{\rm{c}}}
\leftarrow 0$;\\
6:$~~~$ Synchronize thread-specific parameters ${{\bm{\theta
'}}_{\rm{a}}} = {{\bm{\theta }}_{\rm{a}}}$ and  ${{\bm{\theta
'}}_{\rm{c}}} = {{\bm{\theta }}_{\rm{c}}}$;\\
7:$~~~$  Set $t = {t_0}$ and observe a network state  ${s_t}$;\\
8:$~~~$ \textbf{for} $t \le t_{\max }^{{\rm{A3C}}}$ \textbf{do}\\
9:$~~~~~$ Select an action ${a_t}$  based on the policy  $\pi ({a_t}|{s_t};{{\bm{\theta '}}_{\rm{a}}})$;\\
10:$~~~~~$Receive an immediate reward ${r_t}$ and observe a new
reward ${s_{t + 1}}$;\\
11:$~~~~~$$t = t + 1$;\\
12:$~~~$\textbf{end for}\\
13:$~$ Update the received return by\\
$~~~~~~~~~~~$${R_t} = \left\{ \begin{array}{l}
0,\;\;\;\;\;\;\;\;\;\;\;\;\;\;{\rm{if~}}{s_t}\;{\rm{is~a~terminal~state,}}\\
{V^\pi }({s_t};{{\bm{\theta }}_{\rm{c}}}),\;\;{\rm{if~
}}{s_t}\;{\rm{is~a~non - terminal~state}}{\rm{.}}
\end{array} \right.$ \\
14:$~$  \textbf{for} $i = t - 1$ to  ${t_0}$ \textbf{do}\\
15:$~~~$  $R = {r_i} + \gamma {R_t}$;\\
16:$~~~$    Update the actor accumulative gradient $d{{\bm{\theta }}_{\rm{a}}}$ by using
(31);\\
17:$~~~$ Update the critic accumulative gradient $d{{\bm{\theta }}_{\rm{c}}}$  by using (33);\\
18:$~~~$ Perform an asynchronous update of global parameters
${{\bm{\theta }}_{\rm{a}}}$ and ${{\bm{\theta }}_{\rm{c}}}$
based on (35), respectively;\\
19:$~~~$ $T = T + 1$;\\
20:$~~$ \textbf{end for}\\
21:$~$ \textbf{end while}\\
\end{normalsize}
\label{alg_lirnn}
\end{algorithm}
}

\emph{1. The training procedure of A3C-AFL performs in an
asynchronous way:}
\begin{itemize}
\item \textbf{The A3C Procedure:} The A3C algorithm is adopted for device
selection, UAVs placement, and resource management in UAV-enabled
networks, which is presented in \textbf{Algorithm 1}. The detailed
processes of \textbf{Algorithm 1} are shown as follows. 1) Before
training the A3C model, we load the real-world UAV-enabled
network dataset and mobile devices' information, which generate a
simulated environment for the federated learning scenario. 2) At
the $t$-th global counter, the global A3C network parameters
${{\bm{\theta }}_{\rm{a}}}$  and ${{\bm{\theta }}_{\rm{c}}}$ as
well as thread-specific parameters ${{\bm{\theta '}}_{\rm{a}}}$
and ${{\bm{\theta '}}_{\rm{c}}}$ are initialized. 3) Each UAV
server acts as a learning agent to observe a network state ${s_t}$
by interacting with the environment. 4) Each learning agent
selects an action  ${a_t}$ (i.e., device selection, UAVs
placement, subchannel allocation, and power allocation) according
to the policy probability distribution $\pi
({a_t}|{s_t};{{\bm{\theta '}}_{\rm{a}}})$ in the actor network,
and receives an immediate reward  ${r_t}$ as well as a new state
${s_{t + 1}}$ after executing the action  ${a_t}$. 5). The state
value function  ${V^\pi }({s_t};{{\bm{\theta '}}_{\rm{c}}})$ and the
estimation of advantage function $A({s_t},{a_t})$ are updated in the critic
network, and the probability distribution $\pi
({a_t}|{s_t};{{\bm{\theta '}}_{\rm{a}}})$ is updated in actor
network. 6) At each thread counter $i$, the accumulative gradients
of the thread parameters ${{\bm{\theta '}}_{\rm{a}}}$ and ${{\bm{\theta '}}_{\rm{c}}}$ are updated according to (31) and (33),
respectively. 7) The global network collects the parameters
$d{{\bm{\theta }}_{\rm{a}}}$ and
$d{{\bm{\theta }}_{\rm{c}}}$, then perform asynchronous update of the global parameters
${{\bm{\theta }}_{\rm{a}}}$ and ${{\bm{\theta }}_{\rm{c}}}$ by
(35), before broadcasting them to each thread separately. 8) The A3C
algorithm repeats the above training steps until the number of
iterations gets the maximum global shared counter $T_{\max
}^{{\rm{A3C}}}$. Finally, the trained A3C model can be loaded to
perform device selection, UAVs placement, and resource management
for federated learning model updating.

\linespread{1.1}{
\begin{algorithm}[t]
\begin{normalsize}
\caption{\normalsize Asynchronous Federated Learning (AFL)
Algorithm}
1:$~$ \textbf{Input:} The maximum number of communication rounds
${T^{{\rm{AFL}}}}$, client devices set $\mathcal{K}$, number of
local iteration ${E^{{\rm{AFL}}}}$ per communication round, and
learning rate $\eta $;\\
\emph{\textbf{Server process:}} // running at each UAV server\\
2:$~$ \textbf{Input:} Execute joint device selection, UAVs placement, and resource
management by running \textbf{Algorithm 1}. \\
3:$~$  Initializes global model parameters ${{\bf{w}}_n}(0)$ at
each UAV server $n \in \mathcal{N}$;\\
4:$~$ \textbf{for} each global round $t$ $=$ 0, 1, 2, \dots, ${T^{{\rm{AFL}}}}$   \textbf{do}\\
5:$~~~$  \textbf{for} each UAV server  $n$ $=$ 0, 1, 2, \dots, $N$  \textbf{do} in parallel \\
\emph{\textbf{Device process:}} // running at each selected device \\
6:$~~~~~$  \textbf{for} each selected device $k \in {\mathcal{K}_n}$ in parallel \textbf{do}\\
7:$~~~~~~~$   Initialize ${{\bf{w}}_k}(t) = {{\bf{w}}_n}(t)$; \\
8:$~~~~~~~$  \textbf{for} $j$ $=$ 0, 1, 2, \dots, ${E^{{\rm{AFL}}}}$ \textbf{do}\\
9:$~~~~~~~~~$  Sample $i \in {\mathcal{D}_k}$ uniformly at random and update the local parameters  ${{\bf{w}}_k}(t)$ as follows\\
$~~~~~~~~~~~~~$ ${{\bf{w}}_k}(t) = {{\bf{w}}_n}(t - 1) - \eta \nabla {F_k}\left(
{{{\bf{w}}_n}(t - 1)}
\right)$;\\
10:$~~~~~~$ \textbf{end for}\\
11:$~~~~~~$ UAV server $n$ collects the parameters $\{
{{\bf{w}}_k}(t)\} _{k = 1}^{{K_n}}$ from selected
devices, and updates the global parameters\\
$~~~~~~~~~~~~$ ${{\bf{w}}_n}(t + 1) = {{\sum\nolimits_{k \in {\mathcal{K}_n}}
{|{\mathcal{D}_k}|{{\bf{w}}_k}(t)} } \mathord{\left/
 {\vphantom {{\sum\nolimits_{k \in {\mathcal{K}_n}} {|{\mathcal{D}_k}|{{\bf{w}}_k}(t)} } {\left| {{\mathcal{D}_n}} \right|}}} \right.
 \kern-\nulldelimiterspace} {\left| {{\mathcal{D}_n}} \right|}}$ ;\\
12:$~~~~~$ \textbf{end for}\\
13:$~~~$ \textbf{end for}\\
14:$~$ \textbf{Output:} Finalized global FL model parameters ${{\bf{w}}_n}$ of each UAV server  $n \in \mathcal{N}$.\\
\end{normalsize}
\label{alg_lirnn}
\end{algorithm}
}

\item \textbf{The AFL Procedure:} This procedure consists of two phases,
i.e., local training and global aggregation, which is provided in
\textbf{Algorithm 2}. After performing \textbf{Algorithm 1}, the
following procedure is implemented at each global communication
round $t$ for a federated learning system. 1) Each UAV sever $n$
broadcasts its global model parameters ${{\bf{w}}_n}(t - 1)$  to
its associated devices, where the local model parameters at each
participated device $k$ is set as  ${{\bf{w}}_k}(t) =
{{\bf{w}}_n}(t - 1)$, $k \in {\mathcal{K}_n}$. 2) In the $n$-th
UAV-enabled cell, each selected device $k \in {\mathcal{K}_n}$ updates its
local model parameters in an iterative manner according to the
gradient of its loss function  ${F_k}\left( {{{\bm{w}}_n}(t - 1)}
\right)$. At each local iteration $j$, the local parameters
${{\bf{w}}_k}(t)$  are calculated by (20). 3) The ${K_n}$ selected
devices upload their updated local model-parameters $\{
{{\bf{w}}_k}(t)\} _{k = 1}^{{K_n}}$ to its associated UAV server
$n$. 4) Each UAV server $n$ aggregates the uploaded local
model parameters from the ${K_n}$ selected devices, and updates
the global model parameters ${{\bf{w}}_n}(t)$ by (21) before broadcasting
them to the associated devices.
\end{itemize}

\emph{2. Asynchronous Implementation of A3C-AFL:}

We load the trained A3C model (i.e., \textbf{Algorithm 1}) to perform device selection, UAVs placement, and resource management in multi-UAV-enabled wireless networks.  Then, the selected devices perform local
training, and upload their local model-parameters to its associated UAV server over the
allocated uplink subchannels. Each UAV server aggregates the collected local model
parameters, and broadcasts the updated global model parameters to
each associated device in the downlink. It is worth noting that
the multiple agents (i.e., UAV servers) load their trained models (i.e., \textbf{Algorithm 1}) to asynchronously search for different
exploration policies by interacting with their environment. In
addition, the local model training is asynchronously executed
among a range of participated devices, in order to enhance the efficiency
of trained local models. Thus, to improve the federated aggregation efficiency, some model parameters update
process of the associated devices may not be used for the global
aggregation sometimes (i.e., \textbf{Algorithm 2}) .

 \subsection{Complexity and Convergence Analysis}

This subsection provides the computational complexity and
convergence analysis of the proposed A3C algorithm for federated
learning systems.

Let us define $J$ and $V$ as the number of DNN layers of the actor
network and the critic network, respectively. Both the actor
network and the critic network are also two fully connected networks.
Define ${J_j}$ as the number of neurons of the $j$-th layer in the
actor network, and define ${V_i}$ as the number of neurons of the
$i$-th layer in the critic network. In the actor network, the
computational complexity of the $j$-th layer is $O\left( {{J_{j -
1}}{J_j} + {J_j}{J_{j + 1}}} \right)$, and the total computational
complexity with $J$ layers is $O\left( {\sum\nolimits_{j = 2}^{J - 1}
({{J_{j - 1}}{J_j} + {J_j}{J_{j + 1}}}) } \right)$. Similarly, in
the critic network, the total computational complexity with $V$
layers is $O\left( {\sum\nolimits_{i = 2}^{V - 1} ({{V_{i -
1}}{V_i} + {V_i}{V_{i + 1}}}) } \right)$. As the proposed A3C
algorithm is comprised of both the actor network and the critic
network, the computational complexity of each training iteration
is  $O\left( {\sum\nolimits_{j = 2}^{J - 1} ({{J_{j - 1}}{J_j} +
{J_j}{J_{j + 1}}})  + \sum\nolimits_{i = 2}^{V - 1} ({{V_{i -
1}}{V_i} + {V_i}{V_{i + 1}}}) } \right)$. We set that there have $E$
episodes in the training phase, and each episode has $T$ time steps .
Thus, the overall computational complexity of the proposed A3C
algorithm in the training process is  $O\left(
{ET(\sum\nolimits_{j = 2}^{J - 1} ({{J_{j - 1}}{J_j} + {J_j}{J_{j +
1}}})  + \sum\nolimits_{i = 2}^{V - 1} ({{V_{i - 1}}{V_i} +
{V_i}{V_{i + 1}}} ))} \right)$.

\textbf{Theorem 1:}   The A3C algorithm can reach convergence by
using policy evaluation in the critic network and policy
enhancement (the actor network) alternatively, i.e., $\pi
({a_t}|{s_t};{{\bm{\theta }}_{\rm{a}}}) \in \prod $  will converge
to a policy ${\pi ^ * }({a_t}|{s_t};{{\bm{\theta }}_{\rm{a}}})$
which guarantees ${Q^{{\pi ^ * }}}({s_t},{a_t})
> {Q^\pi }({s_t},{a_t})$  for  $\pi ({a_t}|{s_t};{{\bm{\theta }}_{\rm{a}}}) \in \prod $ and  $({a_t},{s_t}) \in \mathcal{S} \times \mathcal{A}$, assuming $|\mathcal{A}| < \infty $. At the same
time, it also requires to satisfy the following conditions [33], [34]: 1)
the learning rates ${\beta _a}(t)$ and ${\beta _c}(t)$ of the
actor network and critic network admit:  $\sum\nolimits_{t = 0}^\infty
{{\beta _a}} (t) = \infty ,\;\sum\nolimits_{t = 0}^\infty  {\beta
_a^2} (t) = \infty $, $\sum\nolimits_{t = 0}^\infty  {{\beta _c}}
(t) = \infty ,\;\sum\nolimits_{t = 0}^\infty  {\beta _c^2} (t) =
\infty $, and   $\mathop {\lim }\limits_{t \to \infty } {{{\beta
_a}(t)} \mathord{\left/
 {\vphantom {{{\beta _a}(t)} {{\beta _c}(t)}}} \right.
 \kern-\nulldelimiterspace} {{\beta _c}(t)}} = 0$; 2) The instantaneous reward ${\rm{Var}}\left\{ {{r_t}} \right\}$ is bounded; 3)
The policy function $\pi (a|s;{{\bm{\theta }}_{\rm{a}}})$ is
continuously differentiable in ${{\bm{\theta }}_{\rm{a}}}$; 4) The
sequence $({a_t},{s_t},{r_t})$ is Independent and identically distributed (i.i.d.), and has uniformly bounded
second moments [33], [34].

\emph{Proof:} See Appendix A.

In the following, we discuss the convergence property of the FL algorithm. Let us define the upper bond of the divergence between the federated loss function  $F({\bf{w}}) $ and the global optimal loss function  $F({{\bf{w}}^ * }) $ as  $\left| {F({\bf{w}}) - F({{\bf{w}}^ * })} \right| $, where  ${{\bf{w}}^ * } $ is the global optimal parameters.

\emph{Definition 1:}  The FL algorithm can achieve the global optimal convergence if it satisfies [10], [11]
\begin{equation}
\begin{split}
 \left| {F({\bf{w}}) - F({{\bf{w}}^ * })} \right| \le \varepsilon ,
\end{split}
\end{equation}
where  $\varepsilon  $ is a small positive constant $\varepsilon  > 0 $.

\textbf{Theorem 2:}  When $F({\bf{w}}) $ is a  $\eta  -  $convex and  $\sigma  -  $smooth function, the upper bond of  $\left[ {F({\bf{w}}) - F({{\bf{w}}^ * })} \right] $ can be expressed
\begin{equation}
\begin{split}
F({\bf{w}}) - F({{\bf{w}}^ * }) \le \varepsilon \left( {F({\bf{w}}(0)) - F({{\bf{w}}^ * })} \right).
\end{split}
\end{equation}

\emph{Proof:} The details of the proof can be seen in [10], [11].

For appropriate selections of the iteration numbers, i.e., the global iterations  ${T^{{\rm{AFL}}}}$ and the local iterations  ${E^{{\rm{AFL}}}}$, the FL algorithm will finally coverage to the global optimality (36), the more proof analysis can be found in [10], [11].

 \section{Simulation Results and Analysis}

In this section, we evaluate the performance of our proposed
A3C-AFL algorithm under different parameter settings. In our
simulations, the performance is evaluated in the Python 3 environment on
a PC with Intel (R) Core(TM) i7-6700 CPU @ 3.40 GHz, 16 RAM, and
the operating system is Windows 10 Ultimate 64 bits. We also
compare the performance of the following different
algorithms/approaches:

\emph{ 1) AFL with device selection:} Proposed asynchronous
federated learning framework with adopting   device selection strategy (i.e., high communication and computation capabilities)
to perform
model aggregation.

\emph{ 2) AFL without device selection:} Proposed asynchronous
federated learning framework with adopting random device selection strategy
to perform model aggregation [17].

\emph{ 3) SFL with device selection:} Synchronous federated
learning framework with adopting device selection strategy
to perform
model aggregation [20].

\emph{ 4) A3C-AFL:} Adopting proposed A3C to perform device
selection, UAVs placement, and resource management in the
asynchronous federated learning framework.

\emph{ 5) A3C-SFL:} Adopting proposed A3C to perform device
selection, UAVs placement, and resource management in the synchronous
federated learning framework.

\emph{ 6) Gradient-AFL:} Adopting gradient-based benchmark to
perform device selection, UAVs placement, and resource management
in the asynchronous federated learning framework.


For our simulations, we consider a multi-UAV-enabled network where
four UAVs are deployed in the sky to support the coverage of a
square area of 400 m $\times$ 400 m. At the beginning, four
UAVs are uniformly located in the sky at the height of 150 m. The device training data of each device $k$
follows uniform distribution [5,10] Mbits, and the CPU computation
capacity of the devices range from 1.0 GHz to 2.0 GHz. The
transmit power of each device is set to be equal to 50 mW, and the
maximum transmit power of each UAV is 150 mW.  The transmit data
size of model parameters is 200 kbits, and the weight parameter is
$\lambda  = 0.4$ in (19). Both the actor network and the critic
network are conducted with deep neural networks (DNNs), and they
have three hidden fully-connected layers. Each of the layers in
the actor network or the critic network contains 256 neurons, 256
neurons, and 128 neurons, respectively. The actor network is
trained with the earning rate 0.0001, and the critic network
is trained with the learning rate 0.001 [33], [34]. The discount factor is
$\gamma  = 0.98$. We evaluate the proposed AFL on the MNIST
dataset [15], [16], and the Convolutional Neural Network (CNN) tool is
used to train the local model. In each global communication round,
FL has one global aggregation and 10 iterations for local
training. The relevant simulation parameters are provided
in Table I.

\subsection{Convergence Comparison of Algorithms}

\begin{figure}
\centering
\includegraphics[width=0.55\columnwidth]{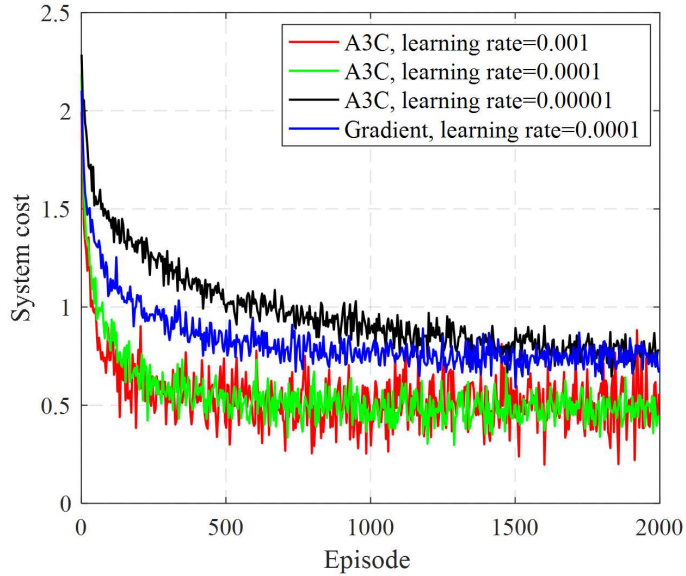}
\caption{{Training convergence of A3C-based learning algorithm in
terms of the system cost.} } \label{fig:Schematic}
\end{figure}

We first evaluate the convergence of the proposed A3C-based
learning algorithm with the different learning rates of the actor network, i.e.,
${\beta _a} = \{ 0.001,0.0001,0.00001\} $, and also compare it
with the gradient-based benchmark algorithm. Note that the system
cost is the objective function in (19), which includes the model
aggregation time and the learning accuracy loss. The learning rate
of DNN plays a key role on the convergence speed and
system cost. From Fig. 3, we can observe that a large value of
learning rate (i.e.,  ${\beta _a} = 0.001$) may cause oscillations
while a small value of the learning rate (i.e., ${\beta _a} =
0.00001$) produces slow convergence. Hence, we select a suitable
learning rate, neither too large nor too small, and the value can
be set around 0.0001 in the training model, which can guarantee
the fast convergence speed and the low system cost. In addition,
it is interesting to note that both the proposed A3C-based
learning algorithm and the gradient-based benchmark algorithm have
the comparable convergence speed, and when the training episode
approximately reaches 500, the system cost gradually converges
despite some fluctuations due to dynamic environment
characteristic and policy exploration. However, our proposed
algorithm achieves the lower system cost than that of the
gradient-based benchmark algorithm. In addition, we also carry out
a test and trials for the critic's learning rate selection, where
the learning rate of the critic network is selected at 0.001 which
has high convergence speed and low system cost [33], [34].

\subsection{ UAVs Placement and Device Selection Evaluation}

Figure 4 captures the 2D deployment of the UAVs (in a horizontal
plane) and selected devices distribution in one time slot, where
a number of $K$=150 devices are randomly located over the
coverage areas of their associated UAVs. UAVs adaptively update
their locations according to the number of associated devices and
devices' distributions, in order to provide the best channel gain
and minimize the communication delay between UAVs and the devices.
It is worth noting that a part of devices with high communication
and computation capabilities (solid dots) are selected to
participate the FL model aggregation, to minimize the aggregation
time and learning accuracy loss, while the remaining low-quality
devices (hollow dots) don't participate in the FL model aggregation
in this time slot.

\begin{figure}
\centering
\includegraphics[width=0.525\columnwidth]{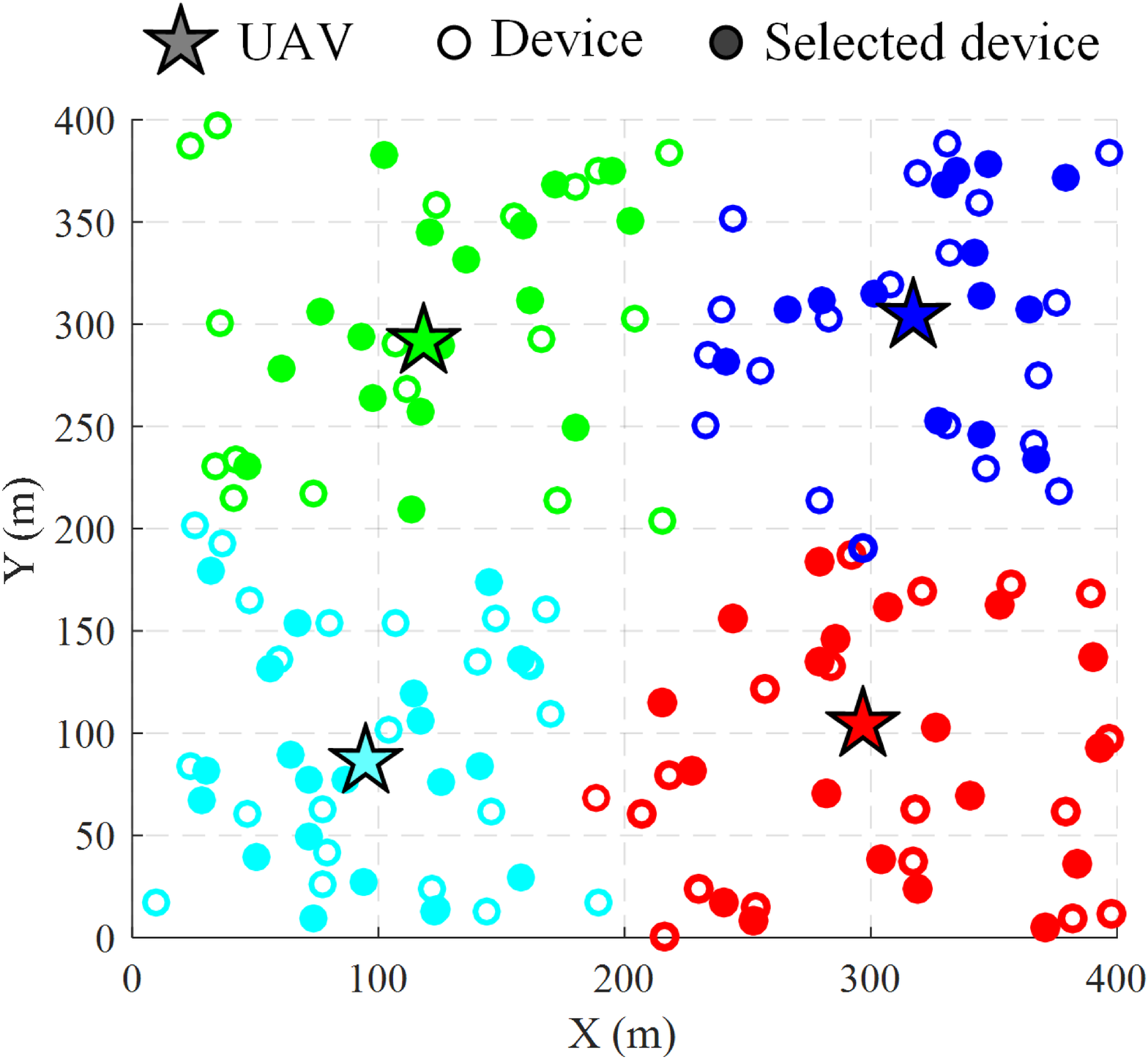}
\caption{{Two dimensional distribution of UAVs, ground devices and
participated devices.} } \label{fig:Schematic}
\end{figure}

\subsection{Accuracy Comparison with Different Global Rounds and Implementation Time}

\begin{figure}
\centering
\includegraphics[width=0.55\columnwidth]{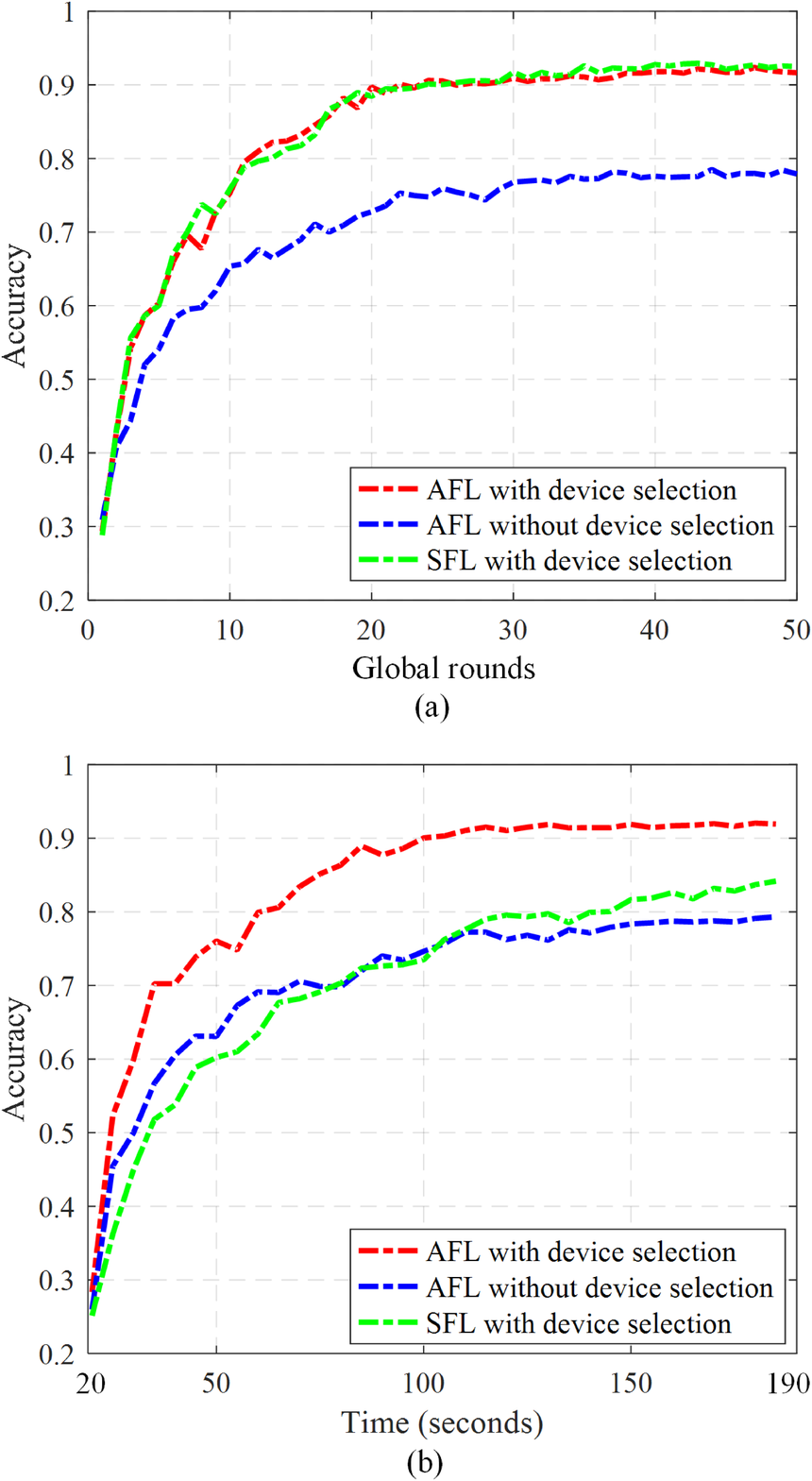}
\caption{{Accuracy comparison versus (a) number of global
communication rounds and (b) wall-clock time.} }
\label{fig:Schematic}
\end{figure}

In Fig. 5, we compare the accuracy performance versus the number
of global rounds and wall-clock time for different FL approaches,
where  several devices are set as low-quality participants. The
low-quality participants are with low communication and
computation capabilities, and even have low-quality training
parameters [14]. From Fig. 5(a), we can observe that both the AFL
and SFL approaches with device selection requires about 25
global rounds to achieve an accuracy of 90.0\%, and both
of them have similar convergence speed and accuracy performance.
However, at each global round, SFL has to wait until all
the selected devices response, while AFL only needs a number of
selected devices' response to move on to the next round which
decreases the aggregation completion time during the learning
process as shown in Fig. 5(b). In addition, both the AFL and SFL
approaches with device selection achieve higher accuracy and
convergence speed than those of the AFL approach without device
selection. The results illustrate that the proposed device
selection scheme can prevent low-quality devices from affecting
the learning accuracy, and enhances the system performance
significantly.

From Fig. 5(b), we can see that the accuracy of all approaches
improves with the increase of the wall-clock time. However, by
comparing different approaches, the proposed AFL approach with device selection
outperform the other two approaches in terms of both the convergence speed and accuracy. The reason lies on the fact
that the AFL approach without device selection may enable the
low-quality devices to participate in the FL model aggregation,
where the low-quality model parameters decrease the overall
accuracy, as well as devices with low communication and computation
capacities need more time to complete the model
aggregation. In addition, the SFL approach with device selection
has to wait for all selected devices to complete their local model
parameters update, among which there may require long computation
time due to low computation capability. Consequently, each global
communication round of the SFL approach with device selection
requires more time to finish model aggregation, and hence its
accuracy performance slowly improves with the increase of
wall-clock time. The results from Fig. 5 demonstrate that the
proposed AFL approach with device selection is capable of improving both
the convergence speed and aggregation accuracy performance.

\subsection{Performance Comparison Versus Number of Devices}

\begin{figure}
\centering
\includegraphics[width=0.55\columnwidth]{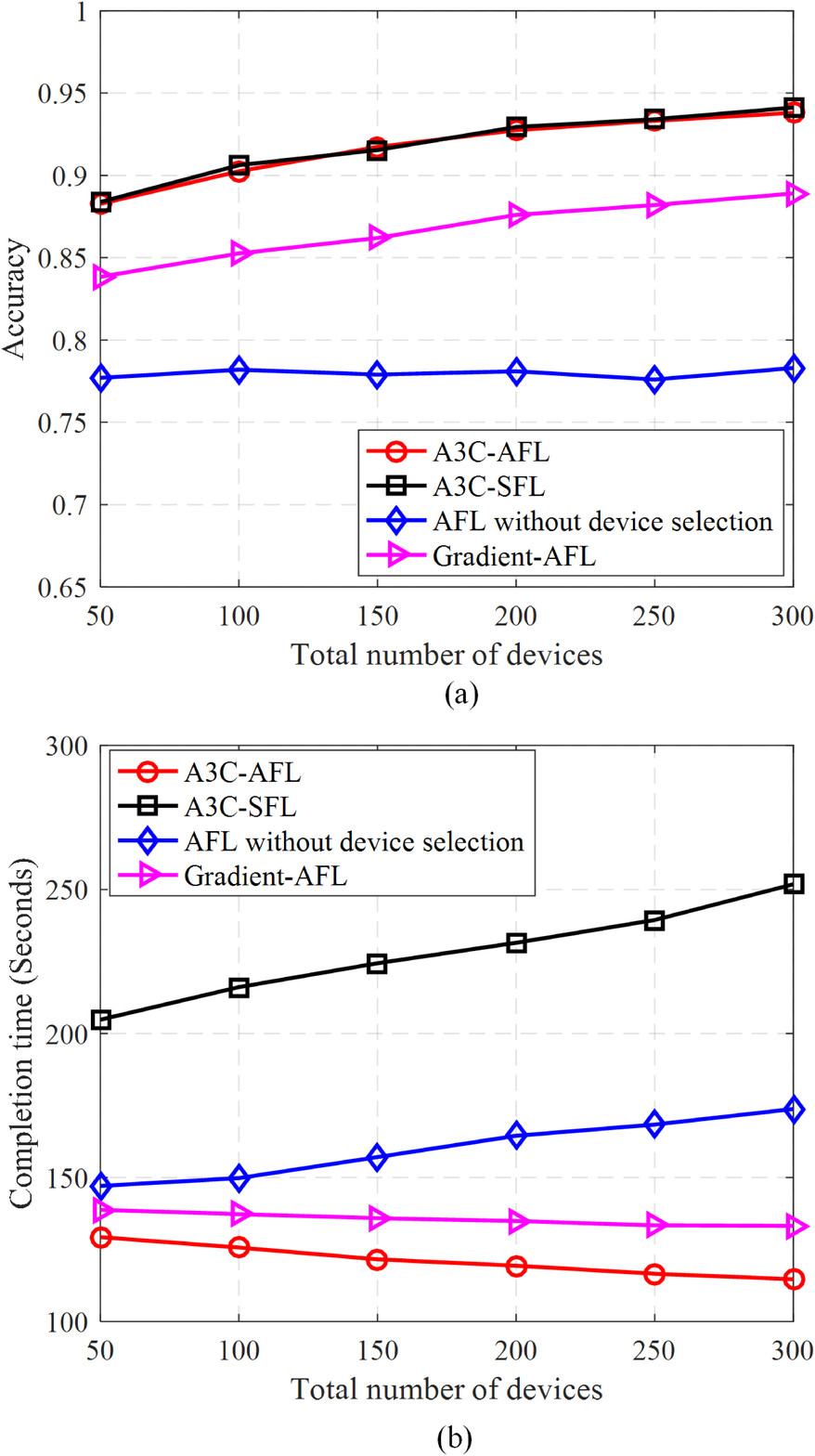}
\caption{{(a) Accuracy and (b) aggregation completion time
comparisons versus total number of devices $K$.} }
\label{fig:Schematic}
\end{figure}

Figure 6 presents the aggregation accuracy and completion time
achieved by different algorithms with various numbers of devices.
As we can see from Fig. 6(a), the proposed A3C-AFL and A3C-SFL
algorithms achieve the superior accuracy compared to the AFL
without device selection algorithm under different numbers of
devices, and the advantage gap becomes large with the increase of
the devices. The reason is that the devices with low-quality model parameters
in the AFL without device algorithm compromise the accuracy after
model aggregation, while the proposed A3C algorithm selects the
devices with high-quality model parameters for model aggregation
which can significantly improve the overall accuracy. Furthermore,
the accuracy of the A3C algorithm increases with the increase of
devices, because the network has the more probability of searching
participated devices having high-quality model parameters to
enhance the aggregation accuracy. However, the accuracy of the AFL
without device selection algorithm maintains at a horizontal level
(i.e., about 78.0\%) with some fluctuation during this process, since
it doesn't keep the low-quality devices from decreasing the
aggregation accuracy. In addition, the accuracy performance of the
gradient-AFL algorithm increases with $K$, but it has a lower
accuracy than that of the proposed A3C-AFL algorithm.

Fig. 6(b) illustrates that the aggregation completion time of the
A3C-SFL and AFL without device selection algorithms increases with
the increase of devices, while the A3C-AFL and gradient-AFL
algorithms decrease slightly during this process. Generally, the
more devices there are, the more completion time is required to learn
the optimal solution for the A3C-SFL algorithm. The reason is that SFL
needs to wait for all selected devices to complete their local
parameters update before aggregating the local models at each
communication round. In addition, for the AFL without device
selection algorithm, as more devices participate in the model
aggregation, the participated devices with low-quality
computation and communication capabilities consume  longer time for
global model aggregation. Even though the gradient-AFL algorithm
prevents low-quality devices from increasing the completion time,
it needs more time to complete the decision making compared to the
A3C-AFL algorithm. The results show that the proposed A3C
algorithm can carry out the better decision making for the device selection, UAVs
placement, and resource management than that of the gradient-based
benchmark.

 \section{Conclusions}

This paper has investigated how to minimize the execution time and learning accuracy loss of privacy-preserving federated learning in multi-UAV-enabled wireless networks. Specifically, an AFL framework was proposed to provide asynchronous distributed computing by enabling model training locally without transmitting raw sensitive data to UAV servers. The device selection strategy was also introduced into the AFL framework to select the mobile  devices  with  high communication  and  computation  capabilities  to improve the learning efficiency and accuracy. Moreover, we also proposed an A3C-based joint device selection, UAVs placement, and resource management algorithm to enhance the learning convergence speed and accuracy. Simulation results have demonstrated that  the proposed AFL framework with A3C-based algorithm outperform the existing solutions in terms of learning  accuracy  and   execution  time under different settings.

  \begin{appendices}
 \section{Proof of Theorem 1}

 For all agents (i.e., UAV servers) in the environment, the network aims to search a joint policy in the coordinated multi-agent RL scenario, where the joint policy can be expressed as
\begin{equation}
\begin{split}
\pi (s{\rm{|}}a) = \left[ \begin{array}{l}
{\pi ^1}({s^1}{\rm{|}}{a^1})\\
\;\;\;\;\; \vdots \\
{\pi ^N}({s^N}{\rm{|}}{a^N})
\end{array} \right] = \left[ \begin{array}{l}
\;\;\;\;\arg \;\mathop {\max }\limits_{{a^1} \in {\mathcal{A}_1}} {Q^1}({s^1}|{a^1})\\
\;\;\;\;\;\;\;\;\;\;\;\;\;\; \vdots \\
\arg \;\mathop {\max }\limits_{{a^N} \in {\mathcal{A}_N}} {Q^N}({s^N}{\rm{|}}{a^N})
\end{array} \right],
\end{split}
\end{equation}
where  ${s^n},{a^n} $, and ${\mathcal{A}_n} $ denote the individual state, action and action space of the $n$-th agent.

As A3C uses the policy evaluation in the critic network and policy enhancement in the actor network alternatively, we adopt the two-time-scale stochastic approximation [33], [34] to provide the convergence proof analysis of A3C. In detail, the convergence of the critic network is first analyzed with the joint policy  $\pi (s|a) $ being fixed. Then, we provide the convergence analysis of the policy parameter ${{\bm{\theta }}_{\rm{a}}} $ upon the convergence of the actor network.

Here, let us define the transition probability of state-action pair as $\mathcal{T}(s',a'|s,a) = \mathcal{T}(s'|s,a)\pi (s'|a') $ and the stationary distribution of MDP as   $D(s,a) = {\rm{diag}}[d(s) \cdot \pi (s|a),s \in \mathcal{S},a \in \mathcal{A}] $, where  $d(s) $ denotes the stationary distribution of the Markov chain induced by policy  $\pi (s|a) $.  The sum cumulative reward of all agents is expressed as  $\bar U(s,a) = \sum\nolimits_{n \in \mathcal{N}} {{U_n}(s,a)}  $, and the overall available cumulative reward set in learning process is defined as  $\hat U(s,a)\bar  = \left[ {\bar U(s,a),s \in \mathcal{S},a \in \mathcal{A}} \right] $. The joint long-term return under the joint policy  $\pi (s|a) $ is expressed as
\begin{equation}
\begin{split}
J({\bm{\theta }}) = \sum\limits_{s \in \mathcal{S}} {d(s)} \sum\limits_{a \in \mathcal{A}} {\pi (s|a)\hat U(s,a)}.
\end{split}
\end{equation}

In addition, the operator  ${W^Q} $ for any action-value function  $Q(s,a;{{\bm{\theta }}_{\rm{c}}}) $ is defined as [33], [34]


 In A3C, the action-value function can be expressed by using the linear functions [33], [34], i.e., $Q(s,a;{{\bm{\theta }}_{\rm{c}}}) = {\bm{\theta }}_{\rm{c}}^T \cdot {\bm{\Psi }}(s,a)$,
where ${\bm{\Psi }}(s,a) = {\left( {\psi {}_1(s,a),\psi {}_2(s,a), \ldots ,\psi {}_{|\mathcal{S}|}(s,a)} \right)^T} $  is the feature vector (basis function vector) at the state $s$. The critic network aims to find a unique solution  ${{\bm{\theta }}_{\rm{c}}} $ by satisfying
\begin{equation}
\begin{split}
{\bm{\Psi }}(s,a) \cdot D(s,a)\left[ {{W^Q}({\bm{\theta }}_{\rm{c}}^T \cdot {\bm{\Psi }}(s,a)) - {\bm{\theta }}_{\rm{c}}^T \cdot {\bm{\Psi }}(s,a)} \right] = 0.
\end{split}
\end{equation}

 As the solution to (41) is a limiting point of the  $TD(0) $ method, and thus we approximate the action-value function  $Q(s,a;{{\bm{\theta }}_{\rm{c}}}) $ instead of the state-value function  $V(s,a;{{\bm{\theta }}_{\rm{c}}})$. This solution can be achieved by minimizing the Mean Square Projected Bellman Error, i.e.,
  \begin{equation}
\begin{split}
\mathop {\min }\limits_{{{\bm{\theta }}_{\rm{c}}}} \left\| {{\bm{\theta }}_{\rm{c}}^T \cdot {\bm{\Psi }}(s,a) - \Pi \left( {W^Q}\left( {{\bm{\theta }}_{\rm{c}}^T \cdot {\bm{\Psi }}(s,a)} \right) \right)} \right\|_{D(s,a)}^2,
\end{split}
\end{equation}
where  $\Pi ( \cdot ) $ denotes the operator that projects a vector to the space, and   $\left\|  \cdot  \right\|_{D(s,a)}^2 $ is the Euclidean norm. In the coordinated scenario, each agent exchanges its decisions with each other, and all of them will achieve a copy of the estimation of the jointly averaged action-value function, i.e.,  ${\bm{\theta }}_{\rm{c}}^n \to {{\bm{\theta }}_{\rm{c}}} $ for all  $n \in \mathcal{N} $. The joint action  $a = \left\langle {{a_1}, \ldots ,{a_N}} \right\rangle  $ of A3C in state $s$ is in a global point that all agents coordinately achieve the sum highest return from the environment, i.e.,  $a = \mathop {\arg \max }\limits_{a \in \mathcal{A}} Q(s,a;{{\bm{\theta }}_{\rm{c}}}) $. In other words, the final critic parameter  ${{\bm{\theta }}_{\rm{c}}} $ is achieved by iteratively minimizing (42) , and the  $Q$ value will converge to the final point  ${Q^ * } $ with probability 1 [33], [34].

To analyze the convergence of the actor network, the advantage function of the $n$-th agent in (29) is rewritten as
\begin{equation}
\begin{split}
A_t^n(s,a) = {\bm{\theta }}_{\rm{c}}^T \cdot {\bm{\Psi }}(s,a) - \sum\limits_{{a^n} \in \mathcal{A}^n} {{\pi ^n}({s_t},{a^n}) \cdot {\bm{\theta }}_{\rm{c}}^T \cdot {\bm{\Psi }}(s,a)} .
\end{split}
\end{equation}

 Using Assumptions 2.2 and 4.1-4.5 [34], for the $n$-th agent, the policy parameter  ${\bm{\theta }}_{\rm{a}}^n $ of the actor network in (35) will converge to a point from the following set of asymptotically stable equilibria of
\begin{equation}
\begin{split}
{\bm{\theta }}_{\rm{a}}^n = {\Gamma ^n}\left[ {{\mathbb{E}_{{s_{t \sim d(s),at \sim \pi }}}}\left( {A_t^n(s,a) \cdot {\nabla _{{\bm{\theta }}_{\rm{a}}^n}}\log {\pi ^n}(s|{a^n})} \right)} \right], {\rm{for}}~n \in \mathcal{N} ,
\end{split}
\end{equation}
where    $\Gamma [ \cdot ] $ denotes an operator that projects any vector onto the compact set. The estimation of the policy gradient  $A_t^n(s,a) \cdot {\nabla _{{\bm{\theta }}_{\rm{a}}^n}}\log {\pi ^n}(s|{a^n}) $ satisfies
\begin{equation}
\begin{split}
\begin{array}{l}
{\mathbb{E}_{{s_{t \sim d(s),at \sim \pi }}}}\left( {A_t^n(s,a) \cdot {\nabla _{{\bm{\theta }}_{\rm{a}}^n}}\log {\pi ^n}(s|{a^n})} \right)\\
 = {\nabla _{{\bm{\theta }}_{\rm{c}}^n}}J({\bm{\theta }}) + {\mathbb{E}_{{s_{t \sim d(s),at \sim \pi }}}}\left( {\left( {Q(s,a;{{\bm{\theta }}_{\rm{c}}}) - {\bm{\theta }}_{\rm{c}}^T \cdot {\bm{\Psi }}(s,a)} \right) \cdot {\nabla _{{\bm{\theta }}_{\rm{a}}^n}}\log {\pi ^n}(s|{a^n})} \right).
\end{array}
\end{split}
\end{equation}

As the linear features here are not limited by the compatible features, we can get the convergence to the stationary point of
 ${\mathbb{E}_{{s_{t \sim d(s),at \sim \pi }}}}\left( {A_t^n(s,a) \cdot {\nabla _{{\bm{\theta }}_{\rm{a}}^n}}\log {\pi ^n}(s|{a^n})} \right) = 0 $ in the set of the policy parameters. In this case, when the long-term averaged return  $J({\bm{\theta }}) $ satisfies   ${\nabla _{{\bm{\theta }}_{\rm{c}}^n}}J({\bm{\theta }}) = 0 $, the error between the approximation function  ${\bm{\theta }}_{\rm{c}}^T \cdot {\bm{\Psi }}(s,a) $ and the action value function  $Q(s,a;{{\bm{\theta }}_{\rm{c}}}) $ is small, i.e.,  $Q(s,a;{{\bm{\theta }}_{\rm{c}}}) - {\bm{\theta }}_{\rm{c}}^T \cdot {\bm{\Psi }}(s,a) \approx 0 $.    Thus, we can achieve the best solution for A3C with general linear function approximation [33], [34]. Due to the update rule in (45) and the coordination nature of the coordinated multi-agent A3C, a joint policy  $\pi  $ implied by multiple policies is updated by each agent, eventually converges to the final point. The more details of the proof can be seen in [33], [34].

\end{appendices}

\end{document}